\documentclass[twocolumn]{aastex62}
\usepackage{natbib, color, amsmath}
\begin{document}

	\title{On the production of He$^+$ of solar origin in the solar wind}
	
	\author{Yeimy J. Rivera}
	\affiliation{Department of Climate and Space Sciences and Engineering, University of Michigan, Ann Arbor MI 48109, USA}
	\author{Enrico Landi} 
	\affiliation{Department of Climate and Space Sciences and Engineering, University of Michigan, Ann Arbor MI 48109, USA}
	\author{Susan T. Lepri}
	\affiliation{Department of Climate and Space Sciences and Engineering, University of Michigan, Ann Arbor MI 48109, USA}
	\author{Jason A. Gilbert}
	\affiliation{Department of Climate and Space Sciences and Engineering, University of Michigan, Ann Arbor MI 48109, USA}

	\begin{abstract}
		
		Solar wind measurements in the heliosphere are predominantly comprised of protons, alphas, and minor elements in a highly ionized state. The majority of low charge states, such as He$^{+}$, measured in situ are often attributed to pick up ions of non-solar origin. However, through inspection of the velocity distribution functions of near Earth measurements, we find a small but significant population of He$^+$ ions in the normal solar wind whose properties indicate that it originated from the Sun and has evolved as part of the normal solar wind.  Current ionization models, largely governed by electron impact and radiative ionization and recombination processes, underestimate this population by several orders of magnitude. Therefore, to reconcile the singly ionized He observed, we investigate recombination of solar He$^{2+}$ through charge exchange with neutrals from circumsolar dust as a possible formation mechanism of solar He$^{+}$. We present an empirical profile of neutrals necessary for charge exchange to become an effective vehicle to recombine He$^{2+}$ to He$^{+}$ such that it meets observational He$^{+}$ values. We find the formation of He$^{+}$ is not only sensitive to the density of neutrals but also to the inner boundary of the neutral distribution encountered along the solar wind path. However, further observational constraints are necessary to confirm that the interaction between solar $\alpha$ particles and dust neutrals is the primary source of the He$^{+}$ observations.
		
	\end{abstract}
	
	
	\section{Introduction} \label{intro}
	
	As solar material is released from the photosphere into the solar wind, it simultaneously experiences a rapid rise in temperature and decrease in density as it travels through the chromosphere and transition region layers and then enters the multimillion degree corona. As it travels through the corona, the density continues to steadily decrease as the temperature evolves more slowly. Once density has decreased enough to effectively stop ionization and recombination processes, the solar wind  ionization status remains constant (e.g. "frozen-in") in the heliosphere. Given the high temperature of the corona, the solar wind ionization state consists of fully ionized H and He along with highly ionized minor ions. Therefore, it is predicted that the solar wind includes very few neutrals or low ionized material as it travels through the interplanetary medium.
	
	Consequently, low-ionized plasma detected by in-situ instruments is not ordinarily attributed to the solar wind, but rather associated with pick-up ions (PUIs; \citealt{Mobius1985, Gloeckler1993}) of interstellar origin or inner source PUIs which are predicted to be formed through the interaction with interplanetary dust \citep{Geiss1994, Geiss1995}. PUIs are usually singly ionized particles that originate as neutrals of non-solar origin such as interstellar neutrals, cometary material \citep{Nordholt2003, Gilbert2015}, interplanetary dust \citep{Gloeckler2000, Grun2001}, or planetary wakes \citep{Russell1981, Grunwaldt1997} that enter or exist in the heliosphere.  Once in the heliosphere, the neutrals can be ionized through photoionization or electron impact ionization, and can also undergo charge exchange with solar wind ions. Once ionized, the charged particles are then swept up by the interplanetary electromagnetic field and travel alongside the ambient solar wind. 
	
   Additionally, low ionized ions can be found to originate from prominence material within coronal mass ejections (CMEs; \citealt{Lepri2010, Gilbert2012}). Prominences, or filaments while observed on the solar disk, are photospheric or chromospheric temperature ($\sim 10^{4}$K) plasma that can be seen as cloud-like structures hovering over the limb of the Sun (see recent reviews by \citealt{Labrosse2010} and \citealt{Parenti2014}). Despite being immersed within the coronal environment, prominences can sustain a low temperature throughout their lifetime where neutral or low ionized material is often observed. Prominences are thought to form within a twisted magnetic field that anchors them to the Sun. However, the magnetic field can become unstable, leading to a large scale eruption that releases the prominence plasma into interplanetary space forming a CME. Low ionized plasma that is able to escape without being further ionized during the eruption, can be observed in the extended solar corona as it travels into the heliosphere \citep{Howard2015c, Wood2016, Ding2017}. Therefore, prominence material can periodically contribute to the low ionized charge states observed near the Earth.
	
	One manner in which PUIs can be differentiated from solar material such as prominence plasma is by examining their velocity distribution functions (VDFs) in phase space. PUI VDFs measured at 1AU do not exhibit a Maxwellian profile typical of the solar wind, but instead are amid thermalization governed by wave-particle interaction \citep{Mobius1985, Gloeckler1993, Drews2016}. The newly ionized neutrals, which now respond to electromagnetic forces, begin to gyrate around the solar wind's magnetic field. In phase space, freshly formed PUIs form a ring distribution surrounding the solar wind ion population with a radius of the solar wind speed. Subsequently, as they evolve as part of the solar wind, the PUIs experience strong pitch angle scattering processes arising from plasma instabilities. This transforms their ring distribution to a spherical shell as the particles converge towards the thermodynamic state of the solar wind.
	
\begin{figure}[]
		\centering
		\includegraphics[width=0.5\textwidth]{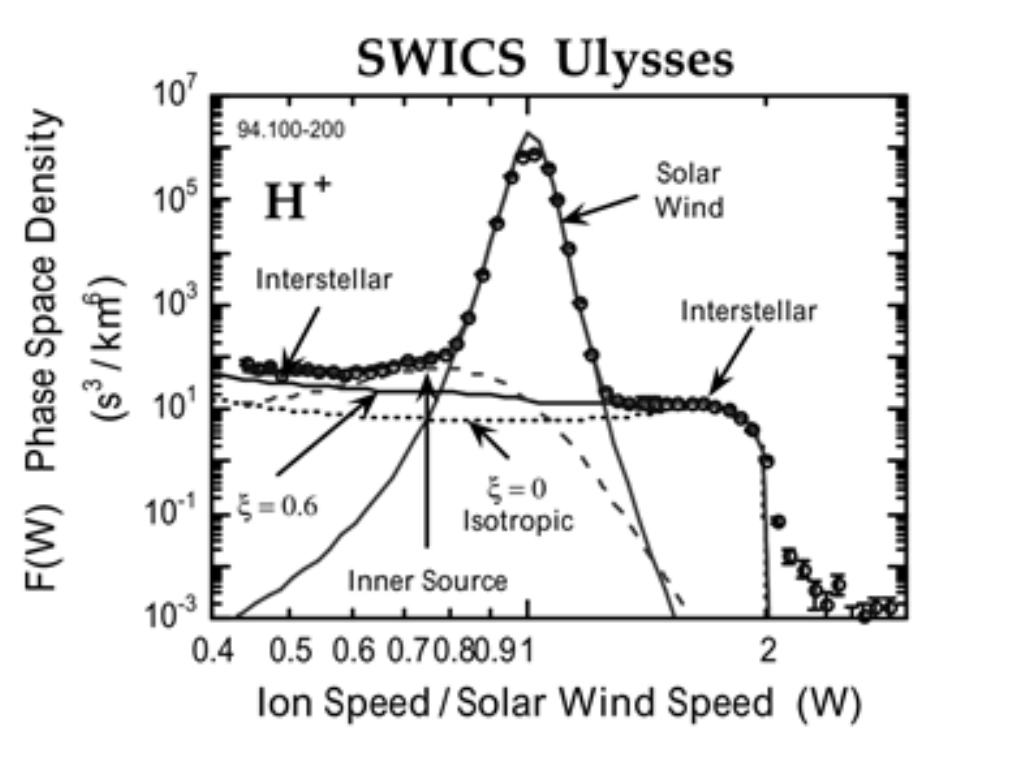}
		\caption{Taken from \cite{Gloeckler1998a}, this figure shows the superposition of the velocity distribution functions of interstellar pick up ions, inner source ions, and solar wind ions in phase space. $\xi$ is the degree of anisotropy in pitch angle scattering from modeling results, see \cite{Vasyliunas1976} and \cite{Thomas1978} for details.}
		\label{fig:PUIVDF}
	\end{figure}

	This is illustrated in measurements of H$^{+}$ from the Solar Wind Ion Composition Spectrometer (SWICS) on Ulysses shown in Figure \ref{fig:PUIVDF}. The figure compares VDFs of the two main PUI populations, the interstellar and inner source PUI, to a typical solar wind distribution. The interstellar PUI VDF uniformly covers velocities below and above the proton VDF, with a sharp drop off at twice the solar wind speed. The inner source PUI VDF, formed closer to the Sun, are farther along the thermalization process as they have had more time to couple to solar wind properties. This distinction in the VDF profiles has become one of the main signatures used in the identification of non-solar material in the heliosphere. 
	
	The helium focusing cone is an example of known interstellar PUIs routinely observed at the distance of the Earth's orbit \citep{Gloeckler2004}. The helium focusing cone is formed from neutral interstellar He that enters the heliosphere in the direction of the interstellar flow. The neutrals penetrate the heliosphere, are gravitationally attracted to the Sun, and subsequently photoionized along their path. This process produces a focused stream of He$^{+}$ PUIs downstream of the Sun. At L1, instruments enter the interstellar He$^{+}$ flow from the helium focusing cone annually during the months of November and December. These PUI measurements have been useful in the characterization of the interstellar medium, in the study of PUI processes in the heliosphere, and in the identification of non-solar material in the solar wind \citep{Gloeckler1998a,  Mobius2004}.
	
	The production of the inner source ions is still debated \citep{Gloeckler1998a, Allegrini2005}; however, many studies propose a connection to the interplanetary dust and, possibly, sungrazing comets \citep{Gloeckler2000, Wimmer-Schweingruber2003, Bzowski2005, Schwadron2008, Mann2010}. Apart from their characteristic non-Maxwellian VDFs, the inner source ion origin is hinted at through studies of their chemical composition. Some studies find agreement between the inner source ion and the solar wind composition suggesting the inner source PUIs originate as part of the solar wind and subsequently formed as result of interaction with interplanetary dust \citep{Brownlee1996, Gloeckler2000, Berger2015}. Conversely, other studies find the elemental composition to differ from solar values \citep{Taut2015}. These discrepancies may be an indication that several processes could be taking part in forming the inner source ion population.
	
	Earlier studies propose that neutrals (H, H$_{2}$, He) outgassed from the interplanetary dust grains can charge exchange with solar wind ions \citep{Banks1971, Fahr1981, Gruntman1996}. In this scenario, the dust neutrals encountered by the solar wind alphas and protons can charge exchange resulting in ionized dust material along with singly ionized and neutral solar material. This is akin to charge exchange in cometary environments \citep{Cravens1997,Bodewits2004, Bodewits2006, SimonWedlund2019}. For the reactions, $\alpha + \text{H}_{2}\rightarrow \text{He}^{+}+\text{H}_2^+$ and $\alpha + \text{H} \rightarrow \text{He}^{+}+\text{H}^+$ where $\alpha$ and He$^{+}$ originate as part of the solar wind, studies find that charge exchange can produce a non-negligible amount of solar He$^{+}$ ions. This process would produce a singly ionized outgassed dust ion along with singly ionized He, from solar alphas, consistent with solar wind properties e.g. a Maxwellian profile narrowly peaked around the proton speed that is characterized by solar wind temperature. 
	
	Previous studies of He$^{+}$ from SWICS on the Advanced Composition Explorer (ACE) investigate the interstellar component of the VDF; however, also find a distribution that peaks at the solar wind speed often attributed to inner source ions \citep{Chen2013}. Our present work systematically analyses He$^{+}$ VDFs measured by ACE/SWICS between $1998-2011$ to understand their source. Our analysis identifies several periods where He$^{+}$ VDFs suggest a solar origin. However, we find current ionization models fail at reproducing the amount of solar wind He$^+$ that is measured, therefore we test the effectiveness of charge exchange with interplanetary dust neutrals as a possible mechanism in the formation of He$^+$.
	
	Results from this study have important implications to our current interpretation of the thermal properties of the plasma which can be derived from ion composition measurements made in the heliosphere through nonequilibrium ionization modeling \citep{Ko1997, Rakowski2007, Rakowski2011, Gruesbeck2011, Gruesbeck2012, Rivera2019a}. In these studies, the charge state distributions are reconstructed from the known ionization and recombination processes that are accounted for in the ionization code, as well as the evolution of the wind plasmas's thermodynamic properties. However, if charge exchange does play a role during the solar wind's radial evolution it may be an important process taking part in shaping the charge state distributions observed in situ.  
	
	In this work, Sections \ref{ACEObs} and \ref{Heporigin} explores the He$^{+}$ measurements from ACE/SWICS and their connection to the Sun, respectively. Section \ref{meth} describes the model for the singly ionized He$^{+}$ population of solar origin formed through charge exchange of solar wind alphas with neutral H$_{2}$ and H from interplanetary dust grains. Sections \ref{results} and \ref{discussion} present the simulation results and discuss neutrals in the vicinity of the Sun, respectively. Finally, Section \ref{conclusion} includes final remarks and further constrains to modeling results. 

	\begin{figure}[t]
		\centering
		\includegraphics[width=0.5\textwidth]{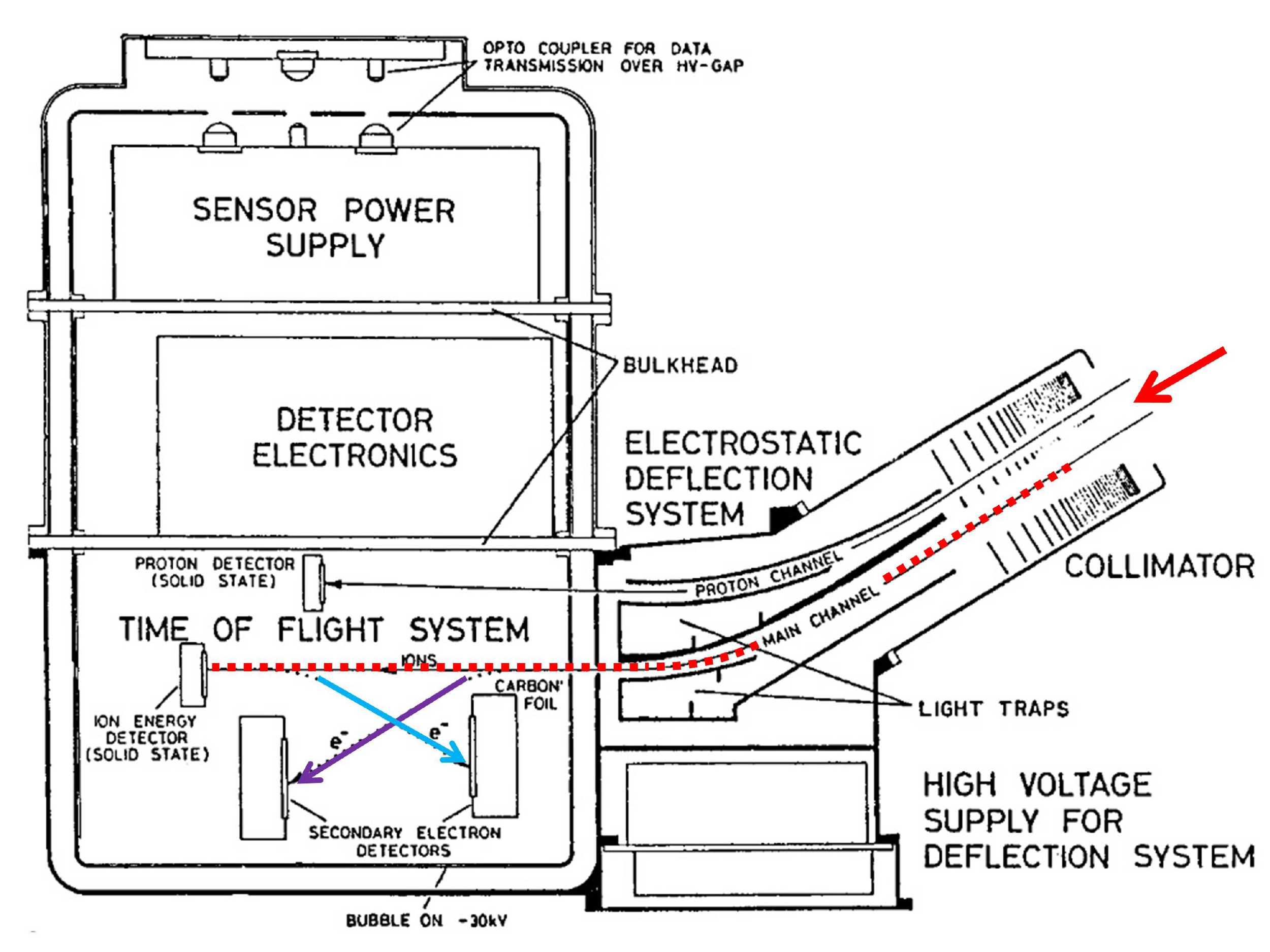}
		\caption{Schematic of the internal compartments of the ACE/SWICS instrument adapted from \cite{Gloeckler1998}. }
		\label{fig:SWICSinstrument}
	\end{figure}
	
	\begin{figure}[t]
		\centering
		\includegraphics[width=0.5\textwidth]{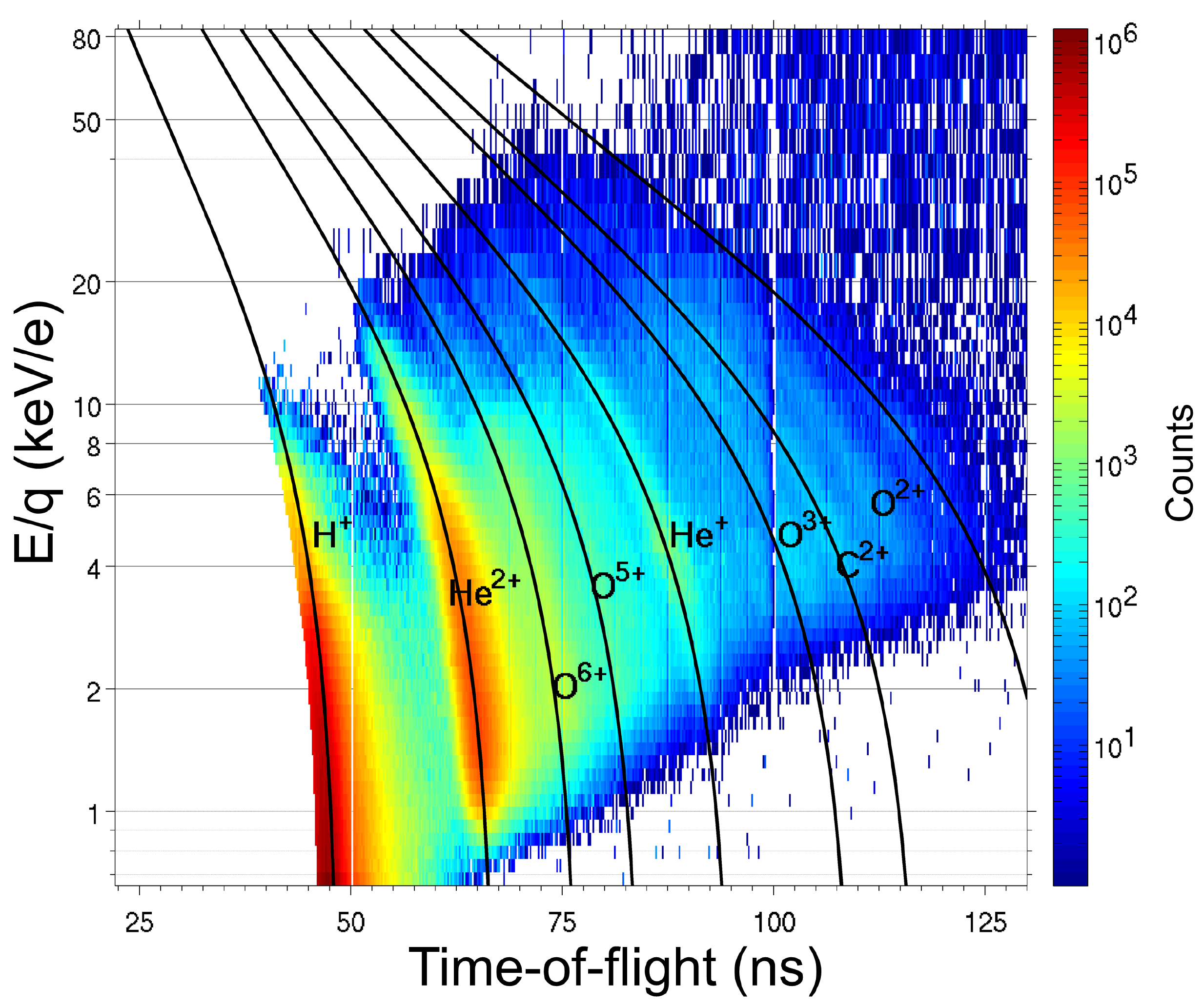}
		\caption{Time-of-flight versus energy-per-charge of ion counts from ACE/SWICS for an accumulation period of 1 January to 31 October 2005.}
		\label{fig:TOFtracks}
	\end{figure}

	\begin{figure*}[t]
	\centering
	\includegraphics[width=\textwidth]{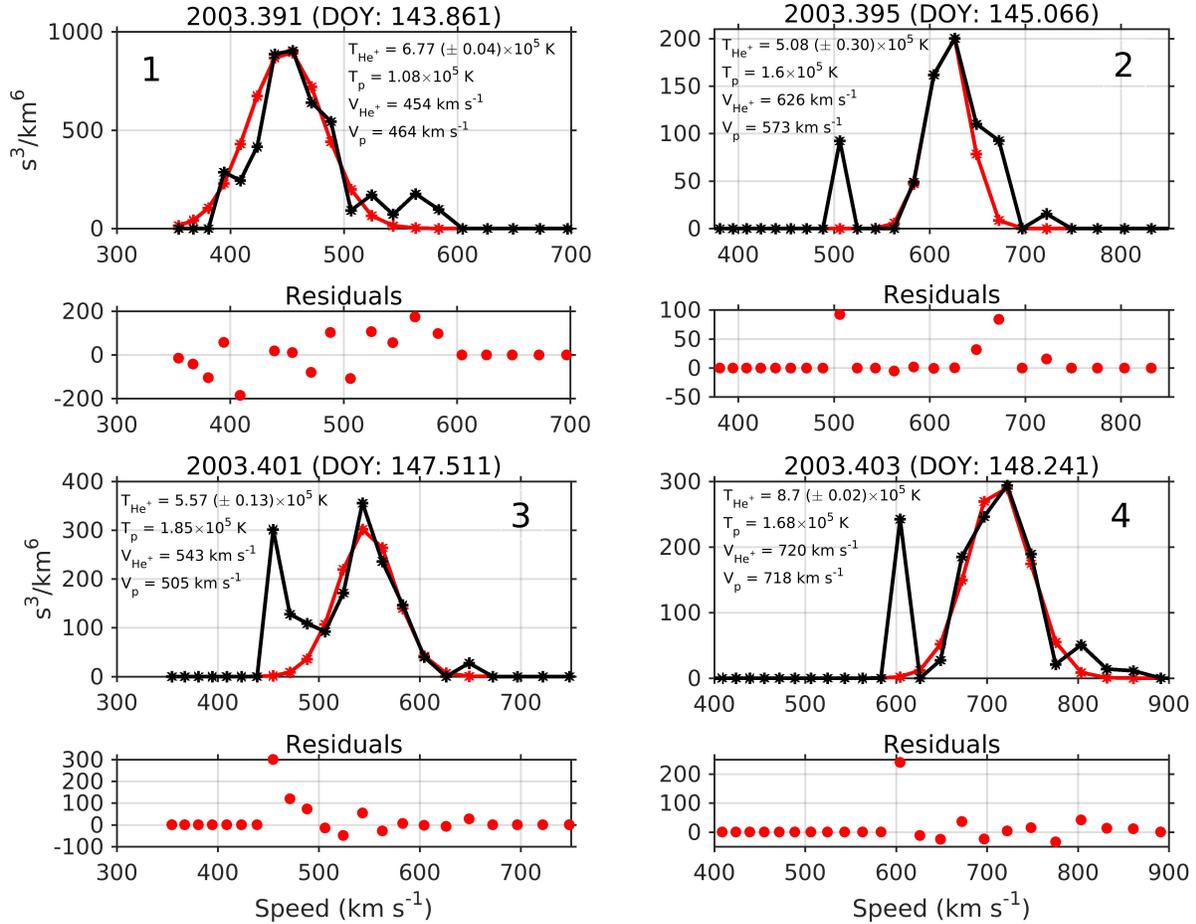}
	\caption{Two panel vertical plots for VDFs of 1 hour accumulation periods (top) and residuals (bottom) in 2003. The black curve are He$^{+}$ measurements and the red curve are Maxwellian fits. We include V$_{p}$, proton speed, V$_{He^+}$, He$^+$ speed, T$_p$, proton temperature, and T$_{He^+}$, He$^+$ temperature where  V$_{p}$ and  T$_{p}$ are 1 hour averages from ACE/SWEPAM and V$_{He^+}$ and T$_{He^+}$ are computed from the Maxwellian fit.}
	\label{fig:VDF}
	\end{figure*}

\begin{table*}[]
	\caption{Summary of VDF properties from Figure \ref{fig:VDF}.}	
	\centering
	\begin{tabular}{c c c c c c} 
		\hline\hline 
		\multicolumn{1}{p{1cm}}{\centering VDF } &
		\multicolumn{1}{p{2cm}}{\centering  T$_{p}(10^{5}$ K)} &
		\multicolumn{1}{p{2.3cm}}{\centering  T$_{He^+} (10^{5}$ K)} &
		\multicolumn{1}{p{2cm}}{\centering  V$_{p}$ (km s$^{-1}$)} &
		\multicolumn{1}{p{2.2cm}}{\centering  V$_{He^{+}}$ (km s$^{-1}$)} &
		\multicolumn{1}{p{2cm}}{\centering He$^{+}$/He$^{2+}$} 
		
		\\
		\hline 
		\\
		1 & 1.08 & $6.77\pm0.04$ & 464 & 454 &  0.03  \\
		
		2 & 1.60 & $5.08\pm0.30$ & 573 & 626 &  0.01  \\
		
		3 & 1.85 & $5.57\pm0.13$ & 505 & 543 &  0.02  \\
		
		4 & 1.68 & $8.70\pm0.02$ & 718 & 720 &  0.06  \\
		
		\\
		\hline
	\end{tabular}
	\label{table:VDFprop}	
\end{table*}

	\begin{figure*}[t]
	\centering
	\includegraphics[width=\textwidth]{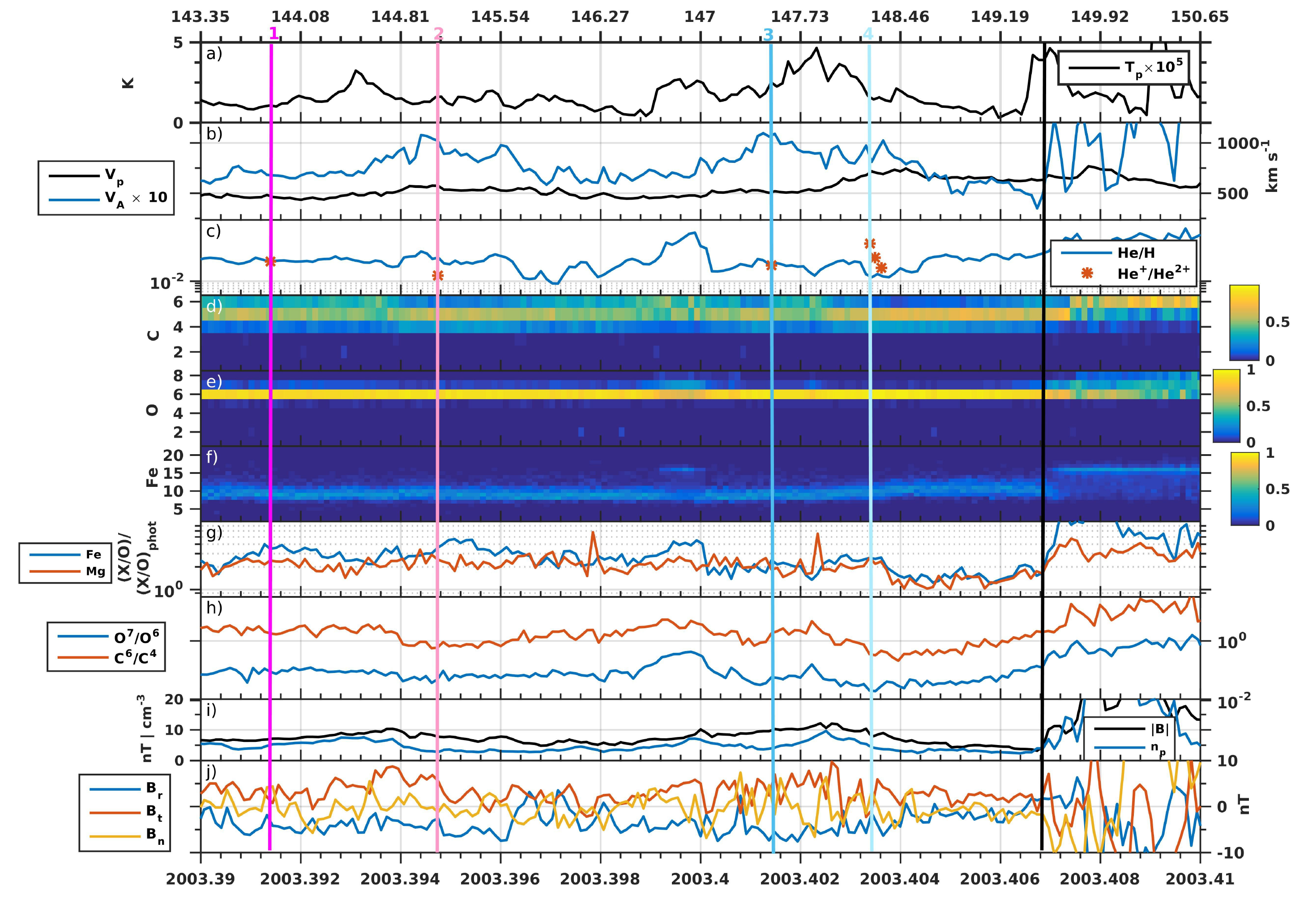}
	\caption{Solar wind properties of the period in 2003 day 143 to 150 from SWEPAM, SWICS, and MAG on ACE. Date in year fraction at the bottom and day fraction at the top. Top to bottom, a) proton temperature, T$_{p}$, b) proton speed, V$_{p}$, and Alfven speed, V$_{A}$, multiplied by 10, c) ratios of He/H and He$^+$/He$^{2+}$ densities, d)-f) relative abundances of C, O, Fe charge states, respectively, g) ratio of Fe/O to photospheric (Fe/O)$_{phot}$ and Mg/O to photospheric (Mg/O)$_{phot}$ using photospheric values from \cite{Asplund2009}, h) O$^{7+}$/O$^{6+}$ and C$^{6+}$/C$^{4+}$, i) magnetic field magnitude, B, and proton density, n$_{p}$, j) radial, B$_{r}$, tangential, B$_{t}$, and normal, B$_{n}$, components of the magnetic field. The colored vertical bars labeled 1-4 correspond to the 1 hour periods from Figure \ref{fig:VDF}. The black vertical solid correspond to beginning of a CME from \cite{Richardson2010}.}
	\label{fig:SolarWindProp}
\end{figure*}

	\begin{figure*}[t]
	\centering
	\includegraphics[width=1\textwidth]{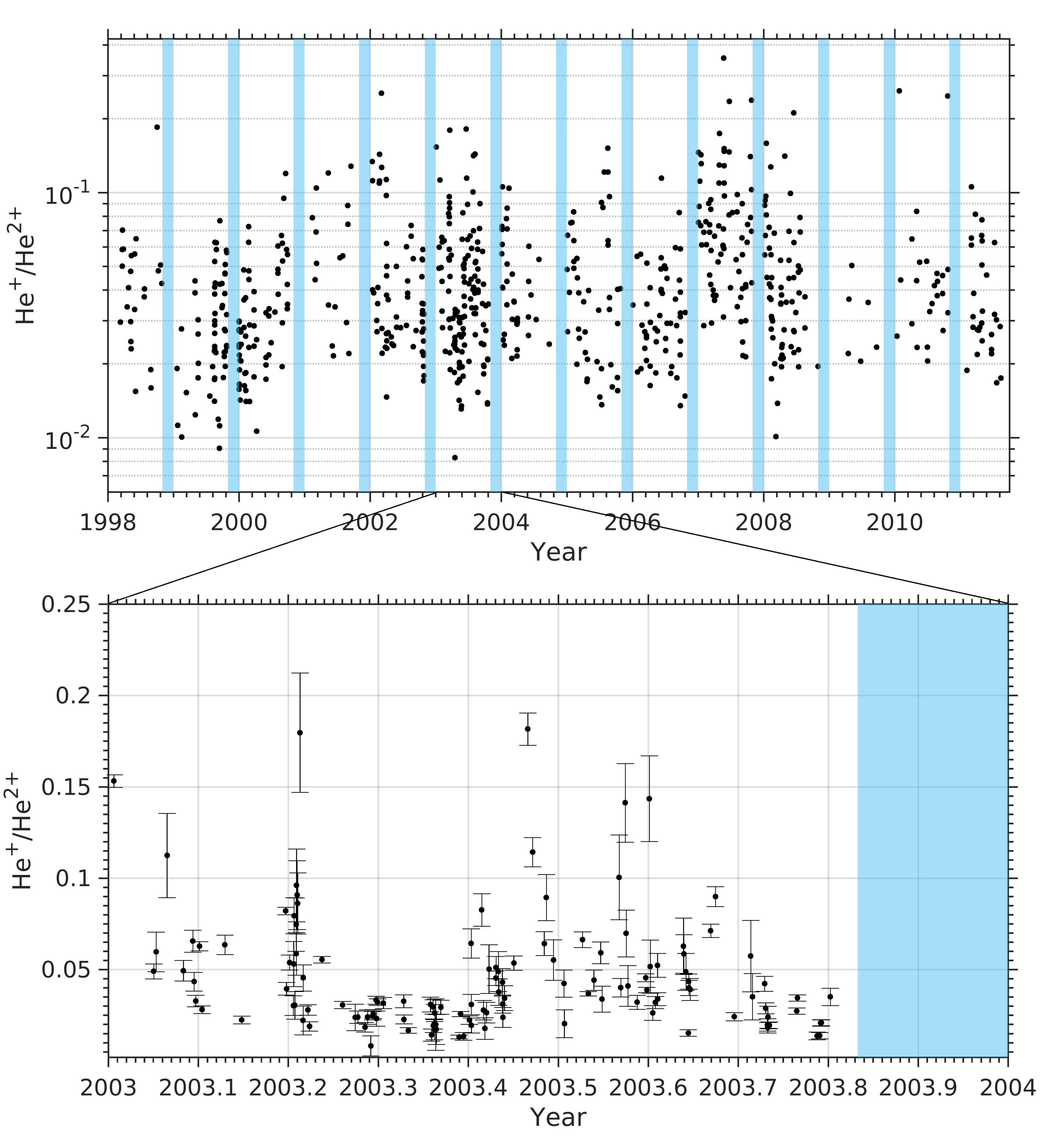}
	\caption{One hour average density ratio of He$^{+}$/He$^{2+}$ from ACE/SWICS for the period between $1998.1-2011.6$ (top) and zoomed into the period of 2003 with corresponding errorbars (bottom). The timeframe highlighted in blue is the Helium focusing cone period, predicted to be between November and December. }
	\label{fig:HeliumTime}
\end{figure*}

	\begin{figure*}[]
	\centering
	\includegraphics[width=\textwidth]{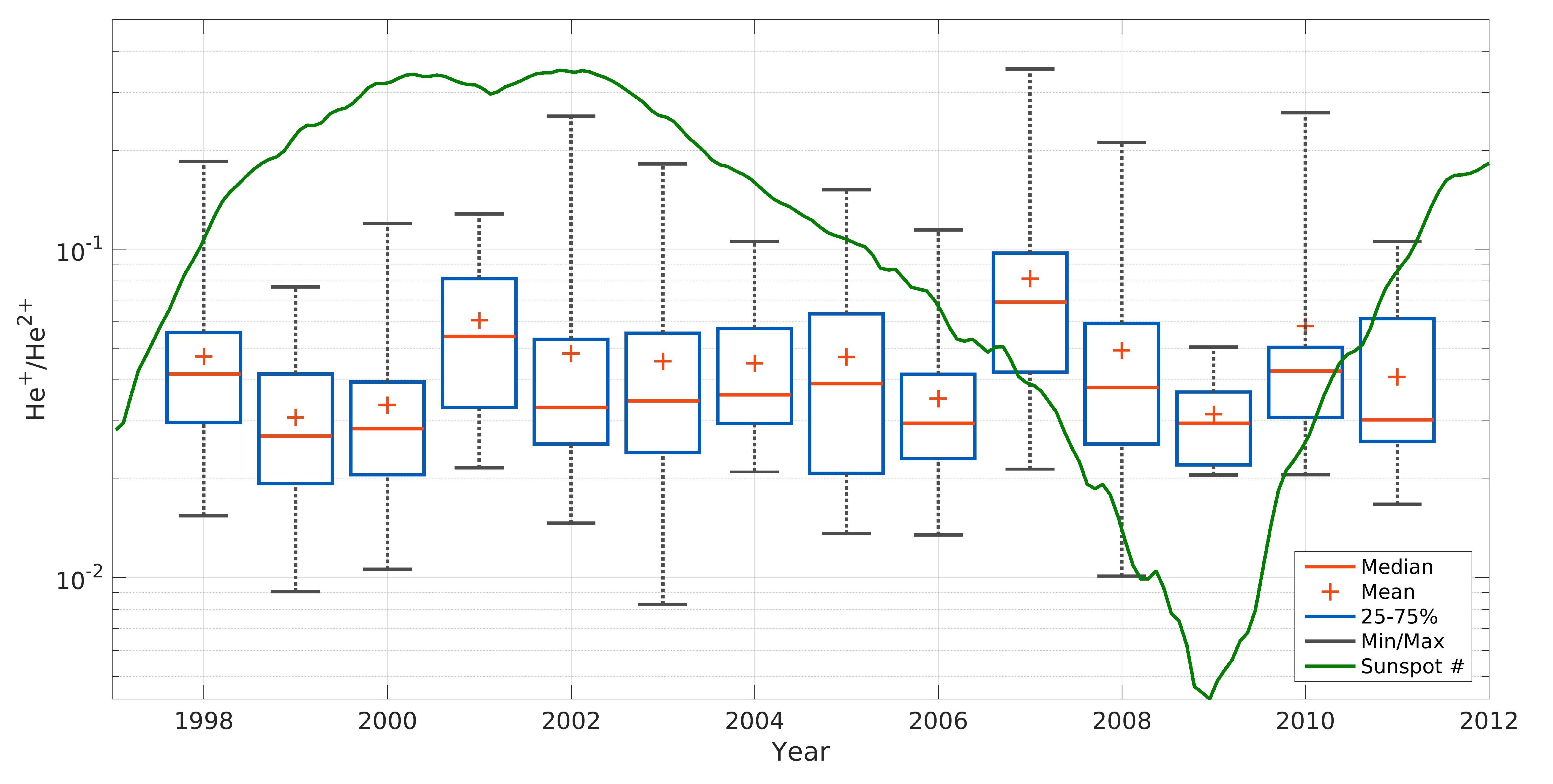}
	\caption{Box plot showing the annual median of He$^+$/He$^{2+}$ represented by the red horizontal bar, annual mean values as a red '+', the blue box shows the range of values within $25-75\%$, and gray whiskers show the range of minimum and maximum annual values. The smoothed sunspot number, normalized to 0.35 for comparison, in plotted in green.}
	\label{fig:AnnualMean}
\end{figure*}
	
	\section{Observations}\label{ACEObs}
	\subsection{Observations from ACE/SWICS}
	He$^{+}$ measurements analyzed are taken from the time-of-flight mass spectrometer, SWICS, aboard ACE between $1998-2011$. The identification of ions with SWICS is determined through a combination of measurements, as fully described in \citep{Gloeckler1994}. The SWICS instrument is shown in Figure \ref{fig:SWICSinstrument}, where the track in dashed red demonstrates the particle's trajectory. 
	
	Initially, the particle enters the instrument through the collimator, immediately deflected by the electrostatic analyzer plates such that only ions of a specific energy-per-charge, $E/Q$, reach the chamber entrance. The selected ions undergo a post-acceleration prior to entering the time-of-flight (TOF) system that increases their energy to exceed the energy threshold of the solid state detector that will measure its energy. Once in the TOF chamber, ions pass through a carbon foil that releases a secondary electron to trigger the start detector, as shown in purple in Figure \ref{fig:SWICSinstrument}. The ion reaches the solid state detector at the end of the compartment which records the ion energy, $E_{meas}$, and releases a second electron to trigger the stop detector, as shown in blue.
	
	From these independent measurements of the TOF, $t$, ion energy, $E_{\text{meas}}$, $E/Q$ from the electric static analyzer, and the post-acceleration voltage, $V$, ions can be unambiguously identified through the following relationships taken from \cite{Gloeckler1998},
	
	\begin{eqnarray}
	M &= 2(t/d)^{2}(E_{\text{meas}}/\alpha) \label{mass}\\
	Q &= \frac{E_{\text{meas}}/\alpha}{V+E/Q-E_{\text{loss}}/Q} \label{charge}\\
	M/Q &= 2(t/d)^2 (V+E/Q-E_{\text{loss}}/Q) \label{masspcharge}\\
	E_{\text{ion}} &= \frac{Q}{(Q/E)} \label{energy} \\
	v_{\text{ion}} &= 438\cdot[(E/Q)/(M/Q)]^{1/2}
	\label{speed}
	\end{eqnarray}
	
	where $d$ is the flight path distance, $V$ is the post-acceleration voltage, $E_{\text{loss}}$ is the energy lost by the ion as it passes through the carbon foil, $\alpha$ is the energy loss due to the solid state detector \citep{Ipavich1978}, and $v_{\text{ion}}$ is the ion speed in km s$^{-1}$.
	
	The particles that produce a concurrent TOF (1) start and (2) stop signal along with the (3) particle's energy measurement are said to produce \textit{triple coincidence measurements}. However, not all ions can meet the minimum energy requirement to trigger the solid state detector. There are a subset of particles that only log a start and stop TOF signal which are known as \textit{double coincidence} counts. This occurs for many singly and low ionized charge states such as the He$^{+}$ measurements which are of interest in present analysis. These ions are rare in triple coincidence counts except for those with high enough energy. Therefore, the double coincidence analysis enables the detailed study of He$^{+}$ that would otherwise be largely unseen through the triple coincidence measurements.
	
	In the double coincidence counts, the instrument records the TOF, $t$, while the $E/Q$ is known from the electrostatic analyzer. Since $E_{\text{meas}}$ is not recorded, we cannot determine Equation \ref{mass}, \ref{charge}, and \ref{energy}, however we can still retrieve $M/Q$ and $v_{\text{ion}}$ from Equation \ref{masspcharge} and \ref{speed}. These two quantities allow us to determine the wind speed and identify the particle, although particles with the same $M/Q$ ratio are blended together making their interpretation difficult.
	 
	Figure \ref{fig:TOFtracks} shows a histogram of double coincidence counts collected between January 1 through October 31 of 2005 plotted as a function of TOF and $E/Q$. We include a selection of modeled tracks (including He$^+$) calculated using Equation \ref{masspcharge} where individual ions are expected to be found along these tracks. Spurious background counts have been removed from the data set using a solar wind speed threshold as discussed in \cite{Gilbert2014}.
	
	A consequence of not having a direct measurement of $E_{\text{meas}}$, is that ions with the same $M/Q$ ratio overlap in the same track, and require specific analysis \citep{Gilbert2012}. Ions from some of the most abundant elements, such as C$^{3+}$ and O$^{4+}$ with $M/Q=4$ are expected to mix with He$^+$ counts in the double coincidence analysis. To ensure the signal extracted from the data set is dominated by He$^+$, we determine the contribution from C$^{3+}$ and O$^{4+}$ during periods where He$^{+}$ is observed in the triple coincidence data set. These instances occur when higher energy He$^+$ ions manage to generate a signal in the solid state detector which allow for He$^{+}$, C$^{3+}$, and O$^{4+}$ to be unambiguously identified. In these cases, we find $n_{He^{+}}/(n_{He^{+}} + n_{C^{3+}} + n_{O^{4+}})$ were $>80\%$ for over 95$\%$ of He$^+$ observations, suggesting that our signal is dominated by He$^{+}$ counts. Additionally, in the triple coincidence counts, C$^{3+}$ and O$^{4+}$ are well sampled while He$^{+}$ is periodically observed and less efficiently measured. Therefore, the He$^+$ counts in the triple coincidence data are likely still underrepresented compared to C$^{3+}$ and O$^{4+}$. Given the He$^{+}$ counts are estimated to be a lower limit provides additional confidence that the He$^{+}$ measurements likely exceed C$^{3+}$ and O$^{4+}$ during the periods observed together. 
	
	As Figure \ref{fig:TOFtracks} illustrates, the He$^{+}$ counts fall along the $M/Q = 4$ TOF track. The points included in this plot are those that fall within $20\%$ of the proton speed. The He$^{+}$ counts cover a similar range and decrease at higher/lower $E/Q$ values along the TOF track similar to common solar wind ions like He$^{2+}$, O$^{6+}$, O$^{5+}$.
	
	\begin{figure}[]
		\centering
		\includegraphics[width=0.5\textwidth]{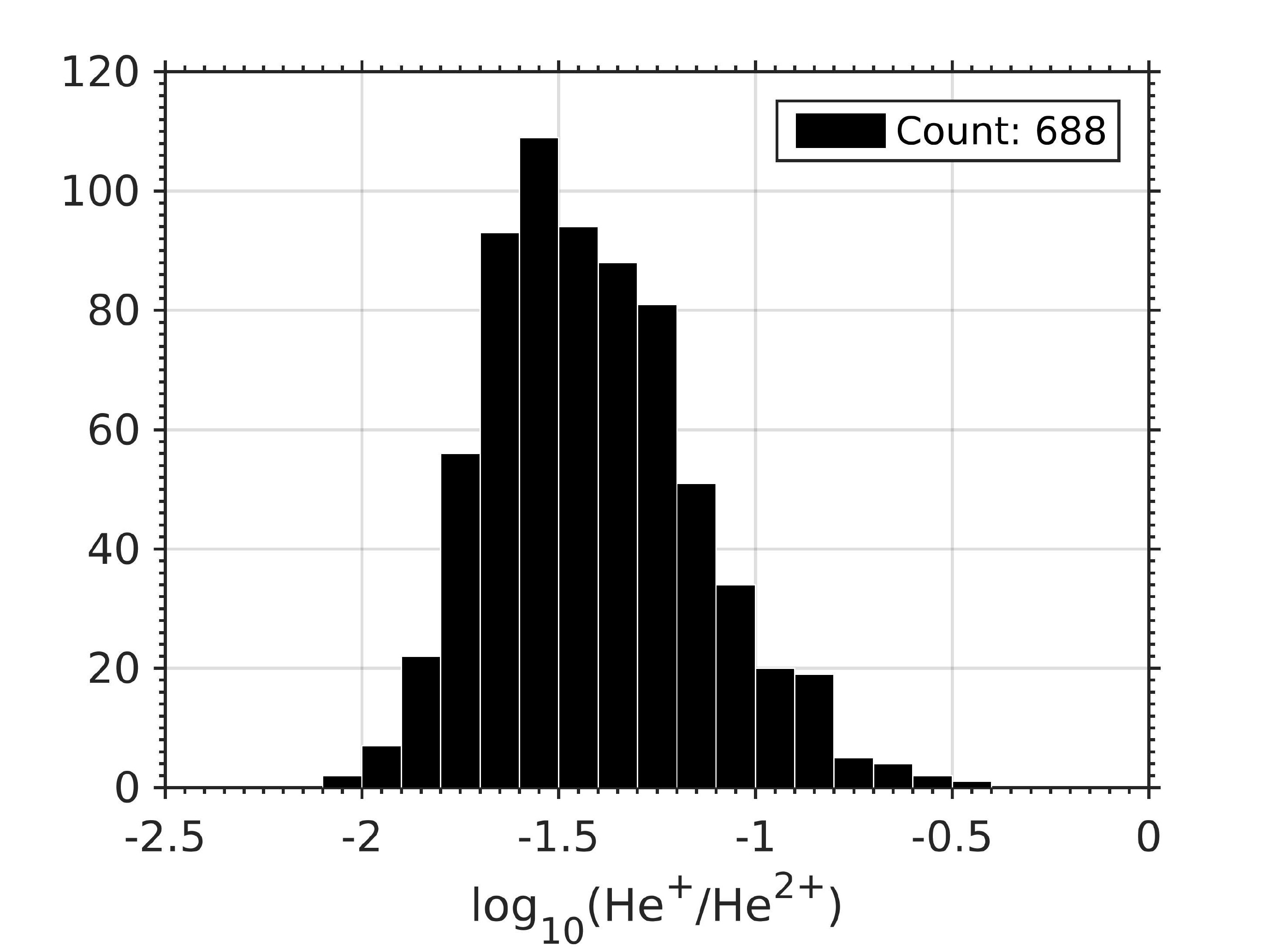}
		\caption{Distributions of He$^{+}$/He$^{2+}$ between $1998-2011$.}
		\label{fig:HeDist}
	\end{figure}
	
	\begin{figure}[]
		\centering
		\includegraphics[width=0.45\textwidth]{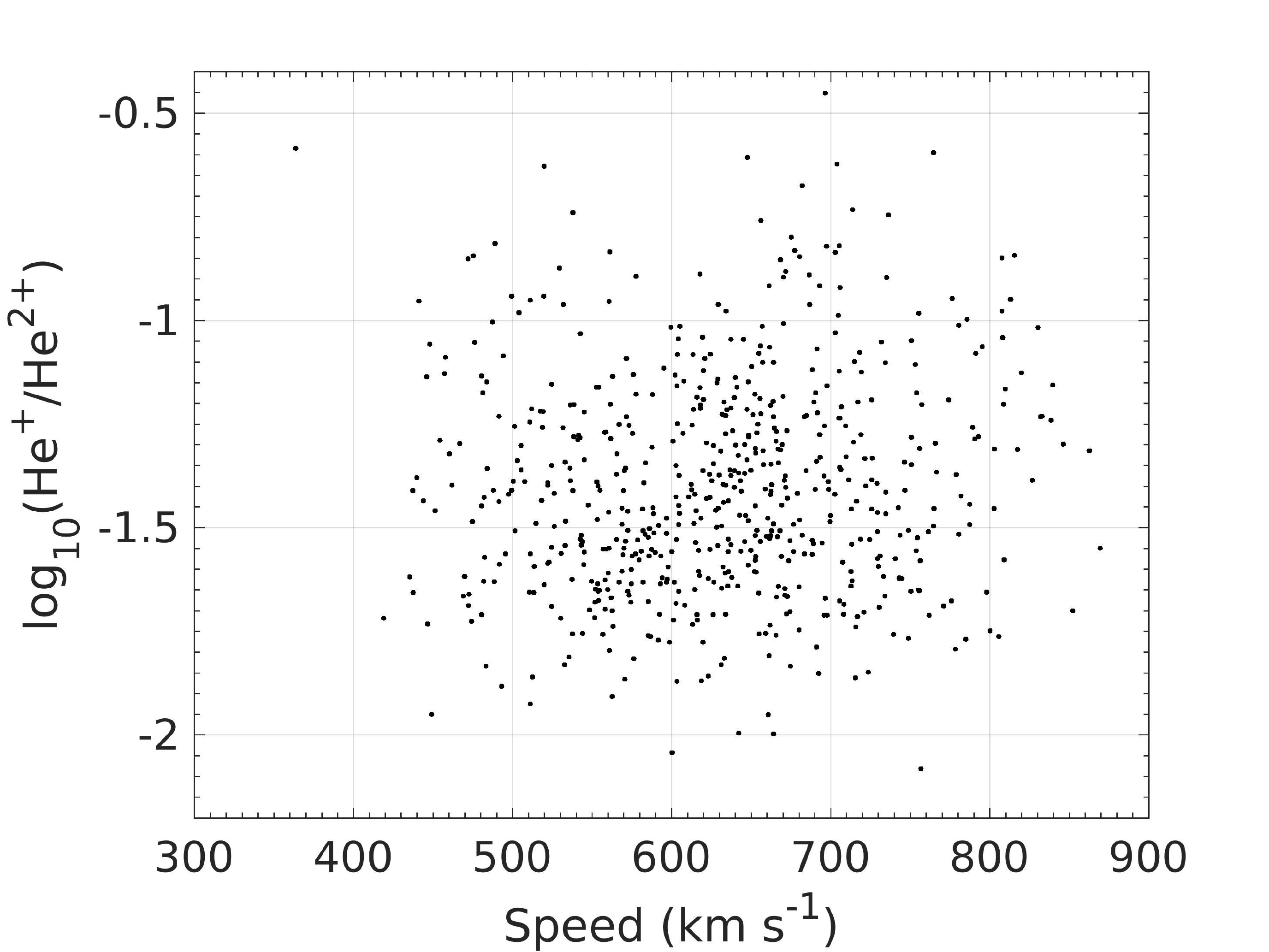}
		\caption{Scatter plot of  He$^{+}$/He$^{2+}$ against solar wind speed.}
		\label{fig:VelScatt}
	\end{figure}

	\subsection{Solar He$^{+}$ measurements}
	We can determine the origin of heliospheric ions through an inspection of the profile formed by their VDFs.  Ions of solar origin resemble Maxwellian distributions while non-solar PUIs are distinguished by their non-thermal profile.
	
	Through inspection of He$^{+}$ VDFs, we identify several periods throughout $1998-2011$ where the He$^+$ VDFs peak within 20$\%$ of the solar wind speed and that are well modeled by a Maxwellian profile. We search the VDFs for a fit to the distribution made up of at least 5 points with relative errors below $30\%$. The measurements identified through this criteria were also manually inspected to ensure the VDFs were well fitted.
	
	As examples, we present sample VDFs in units of s$^3$ km$^{-6}$ for several 1 hr accumulation periods in 2003, shown in Figure \ref{fig:VDF}. The figure includes a top plot where measurements are in black and a Maxwellian fit is in red, along with a plot of residuals between the fit and measurements plotted below. The counts in panels 2, 3, 4 outside of the Maxwellian fits that produce large residuals, such as at 500 km s$^{-1}$ in panel 2, are counts that are not part of the solar He$^{+}$ distribution and possibly associated with the PUI distribution as shown in Figure \ref{fig:PUIVDF}. Additionally, we include the solar wind properties during the four periods from Figure \ref{fig:VDF} in Figure \ref{fig:SolarWindProp}. Figure \ref{fig:SolarWindProp} includes a multi-panel plot between day $\sim143-150$ in 2003 where the four colored vertical lines labeled 1 through 4 correspond to the labels in the four 1 hour VDFs. The black solid line denotes the boundary of an interplanetary CME identified in \cite{Richardson2010}.
	
	In addition, the solar wind properties for the four periods are summarized in Table \ref{table:VDFprop}. We include the proton temperature, T$_{p}$, He$^+$ temperature, T$_{He^+}$, mean proton speed, V$_{p}$, and He$^{+}$ speed, V$_{He^+}$. The proton speed and temperature are 1 hour averages from Solar Wind Electron, Proton, and Alpha Monitor (SWEPAM; \citealt{McComas1998}) on ACE and the He$^{+}$ speed and temperature are computed from the Maxwellian fit. 
	
	Figure \ref{fig:HeliumTime} shows a time series of all measurements that meet the VDF criteria. The top plot is the He$^{+}$/He$^{2+}$ density ratios between $1998.1-2011.63$ and the bottom plot is zoomed into 2003 with associated errorbars. We note that values of He$^+$/He$^{2+}$ are used instead of an absolute abundance of He$^{+}$ to illustrate enhancements in He$^{+}$ densities compared to the alphas, as well as, to easily compare with the results from the ionization code discussed in Section \ref{meth}. We have removed measurements within CMEs and the helium focusing cone to capture He$^{+}$ in ambient solar wind. The measurements within CMEs and the focusing cone made up approximately $16\%$ and $15\%$ of the total points, respectively. We find that the remaining He$^{+}$/He$^{2+}$ measurements are observed throughout individual years and the solar cycle with values ranging an order of magnitude apart.
	
	A summary of these values is shown in Figure \ref{fig:AnnualMean}. The figure includes the He$^{+}$/He$^{2+}$ annual median denoted as the red bar inside each box, annual mean as a red '+', a blue box showing values within $25-75\%$, and gray whiskers extending between the annual minimum and maximum values. We note that 1998 and 2011 were partially measured. We also include the 13-month smoothed monthly sunspot number\footnote{http://www.sidc.be/silso/datafiles} in green, normalized to 0.35 for a more convenient comparison with He$^{+}$/He$^{2+}$ values. Generally, we find the median values of the He$^{+}$/He$^{2+}$ ratio range between 0.02 and 0.06, with the exception of an elevated value during 2007; however, there is no clear solar cycle dependence observed.

 	Figure \ref{fig:HeDist} shows the distribution for all He$^{+}$/He$^{2+}$ values. The distribution peaks between $2.5 - 3.2 \times 10^{-2}$ and range between $8.0\times10^{-3}$ to $4.0\times10^{-1}$. These values are larger that those previously reported from in situ measurements of Vela which measured He$^{+}$/He$^{2+}$ to be on the order of $\sim 10^{-3}$ \citep{Hundhausen1968}. 
 	
 	When plotting the He$^{+}$/He$^{2+}$ values against solar wind speed, as shown in Figure \ref{fig:VelScatt}, the He$^{+}$/He$^{2+}$ ratio is uncorrelated to speed, however appears predominantly in wind traveling faster than 400 km s$^{-1}$. This may indicate the process producing He$^{+}$ becomes effective for plasma accelerated to a solar wind speed above this value but has no strong dependence with increasing speed. In addition, given that He$^{+}$ appears independent of speed, it suggests the production of He$^{+}$ is not dependent on the mechanism accelerating the solar wind.

	\begin{figure}[]
		\centering
		\includegraphics[width=0.45\textwidth]{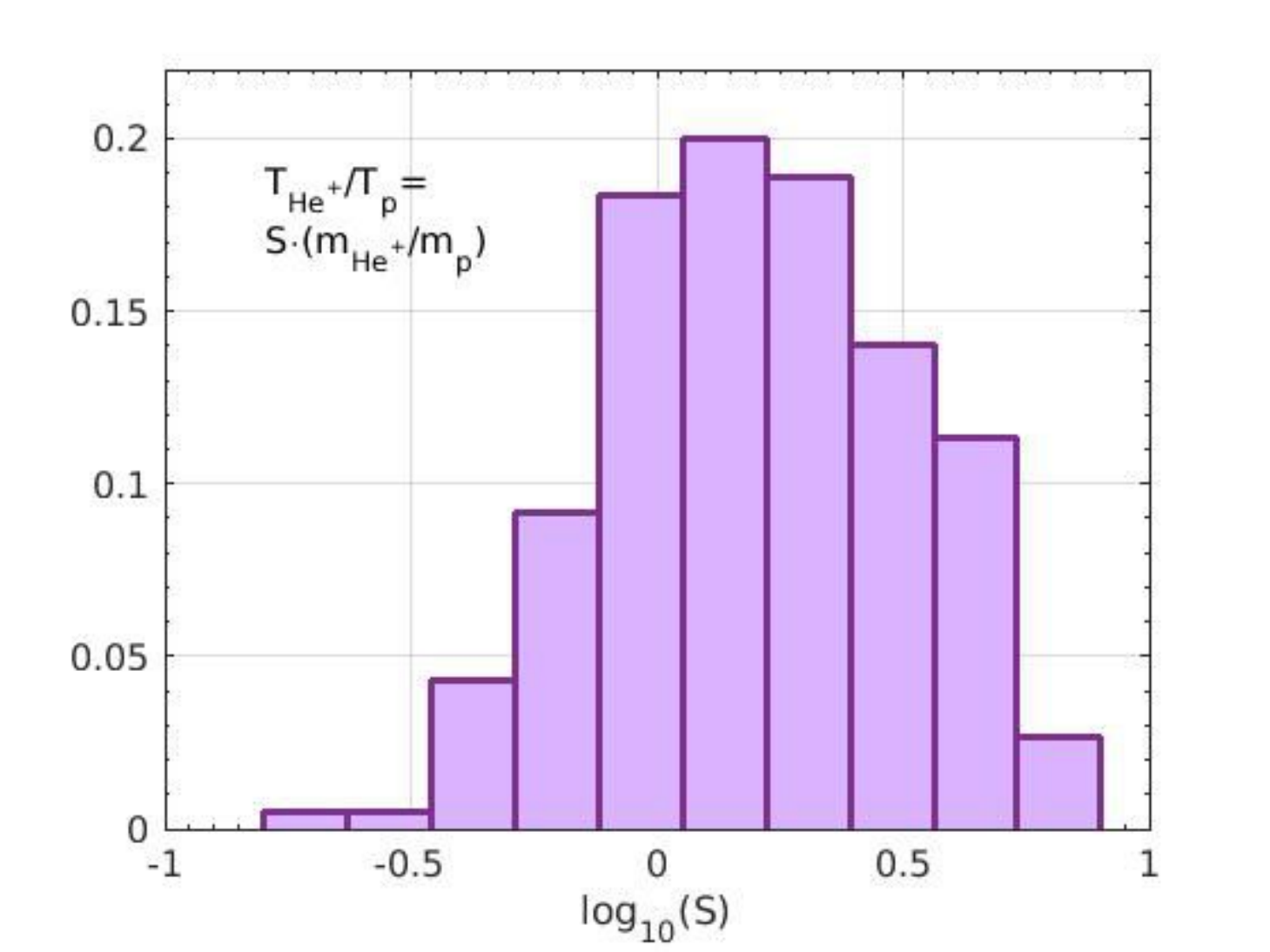}
		\caption{Normalized distribution of S, the proportionally constant in T$_{He^+}$/T$_{p} = S\cdot$(m$_{He^+}$/m$_{p}$).}
		\label{fig:DistTemp}
	\end{figure}

	\begin{figure}[]
	\centering
	\includegraphics[width=0.4\textwidth]{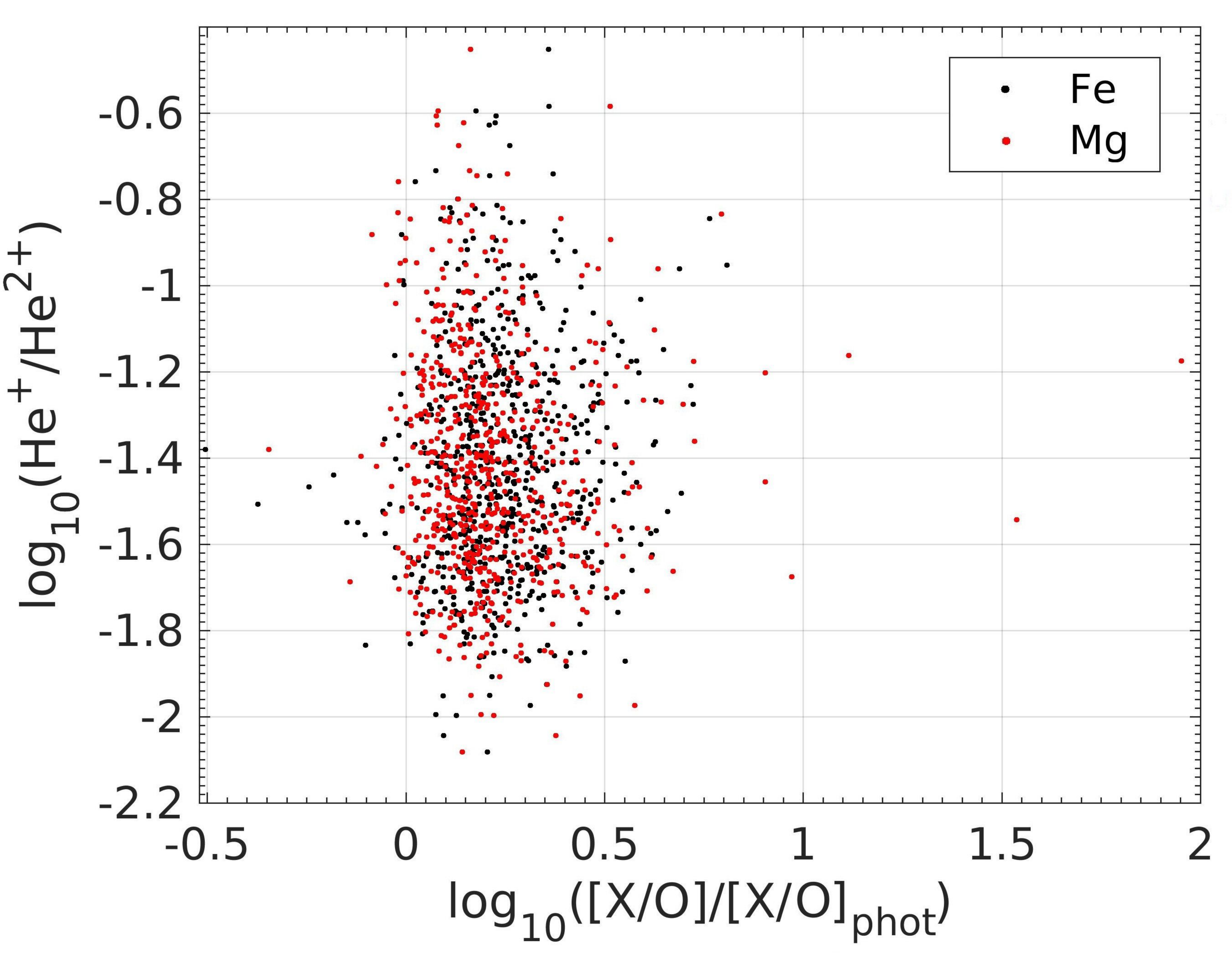}
	\caption{Scatter plot of He$^+$/He$^{2+}$ versus (X/O)/(X/O)$_{phot}$ where $X=$ Fe and Mg.}
	\label{fig:CompScatt}
	\end{figure}

	Furthermore, the temperature values computed for He$^{+}$ coincide with previous measurements from ACE/SWICS which find a mass proportional and super mass proportional relationship for a large number of minor ion temperatures in the solar wind \citep{Tracy2015, Tracy2016}. Collectively for all ions, they find the following relationship, T$_{ion}$/T$_{p} = S\cdot$(m$_{ion}$/m$_{p}$), for $S=1.35\pm0.02$, where T$_{ion}$ and T$_{p}$ are the ion and proton temperature, respectively, and m$_{ion}$ and m$_{p}$ are the ion and proton mass, respectively. A value of $S=1$ describes a mass proportional relationship while $S>1$ would suggest a super mass proportional relationship. Although, the relationship is sensitive to the solar wind's collisional age, A$_{C}$, as described in \cite{Tracy2016}. The proportionality describes ions within low A$_{C}$ and is gradually erased for ions within plasma of higher A$_{C}$ as ions become increasingly coupled to the proton temperature. For the VDFs in Figure \ref{fig:VDF}, we compare the He$^{+}$ temperatures within the different solar wind periods, 1) T$_{He^+}$/T$_{p} = 1.57$ m$_{He^+}$/m$_{p}$, 2) T$_{He^+}$/T$_{p} = 0.8$ m$_{He^+}$/m$_{p}$, 3) T$_{He^+}$/T${p} = 0.75$ m$_{He^+}$/m$_{p}$, and 4) T$_{He^+}$/T$_{p} = 1.30$ m$_{He^+}$/m$_{p}$. The normalized distribution of all values of $S$ are presented in Figure \ref{fig:DistTemp}. We find a relatively low A$_C$ for the He$^{+}$ values, between $8.1\times10^{-3}-1.7\times10^{-2}$. The median $\pm$ median absolute deviation value, as was computed in \cite{Tracy2016}, is $S=1.5\pm0.5$ which is consistent with $S=1.35\pm0.02$ from \cite{Tracy2016}. T$_{He^{+}}$ reflects ion temperatures typical at 1AU, further suggesting the He$^+$ measurements undergo a similar thermal evolution to the solar wind, as well as, are independent of the mechanism heating the solar wind.
	
	When examining the chemical composition of the plasma corresponding to He$^{+}/$He$^{2+}$ measurements, we find the solar He$^{+}$ is independent of solar wind type. The well known First Ionization Potential (FIP) effect has been shown to result in a range of elemental composition between different parts of the Sun, notably between the photosphere and other structures \citep{Feldman1993, Feldman1998, Raymond1997, Raymond1999, Landi2015a, Parenti2019}. This makes the composition measured in the heliosphere a key indicator to the solar wind's origin at the Sun. Measurements of low FIP Fe and Mg to high FIP O compared to the ratio of its photospheric value, (Fe/O)/(Fe/O)$_{phot}$ and (Mg/O)/(Mg/O)$_{phot}$, are commonly used to characterize solar wind streams \citep{vonSteiger2000, Schwadron2002, Zurbuchen2012}. Figure \ref{fig:CompScatt} shows a scatter plot of compositional values for all He$^{+}$/He$^{2+}$ values. The compositional values are well distributed between photospheric, value of 1, to FIP enhanced solar wind, $>1$, showing no preferred solar wind type. This further indicates that the He$^{+}$ is likely independent of the solar wind's birthplace, which is largely governed by the properties discussed, but rather a processes occurring after it leaves the Sun.
	
	\section{Genesis of solar He$^+$ ions}\label{Heporigin}
\begin{figure}[]
		\centering
		\includegraphics[width=0.45\textwidth]{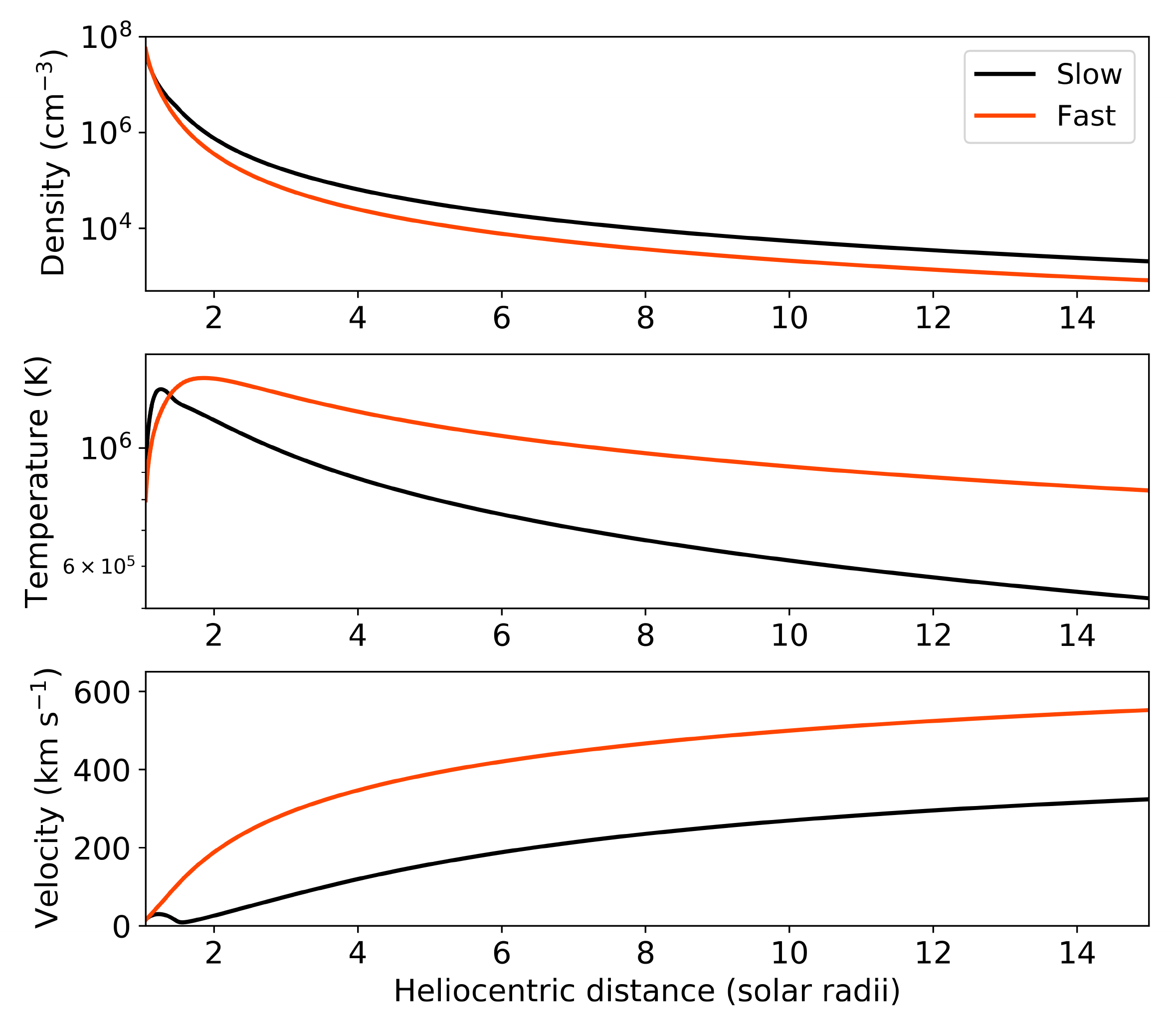}
		\caption{Solar wind thermodynamic evolution for slow and fast wind streams derived in \cite{Cranmer2007}.}
		\label{fig:SolarWindProfiles}
	\end{figure}

\begin{figure}[]
		\centering
		\includegraphics[width=0.45\textwidth]{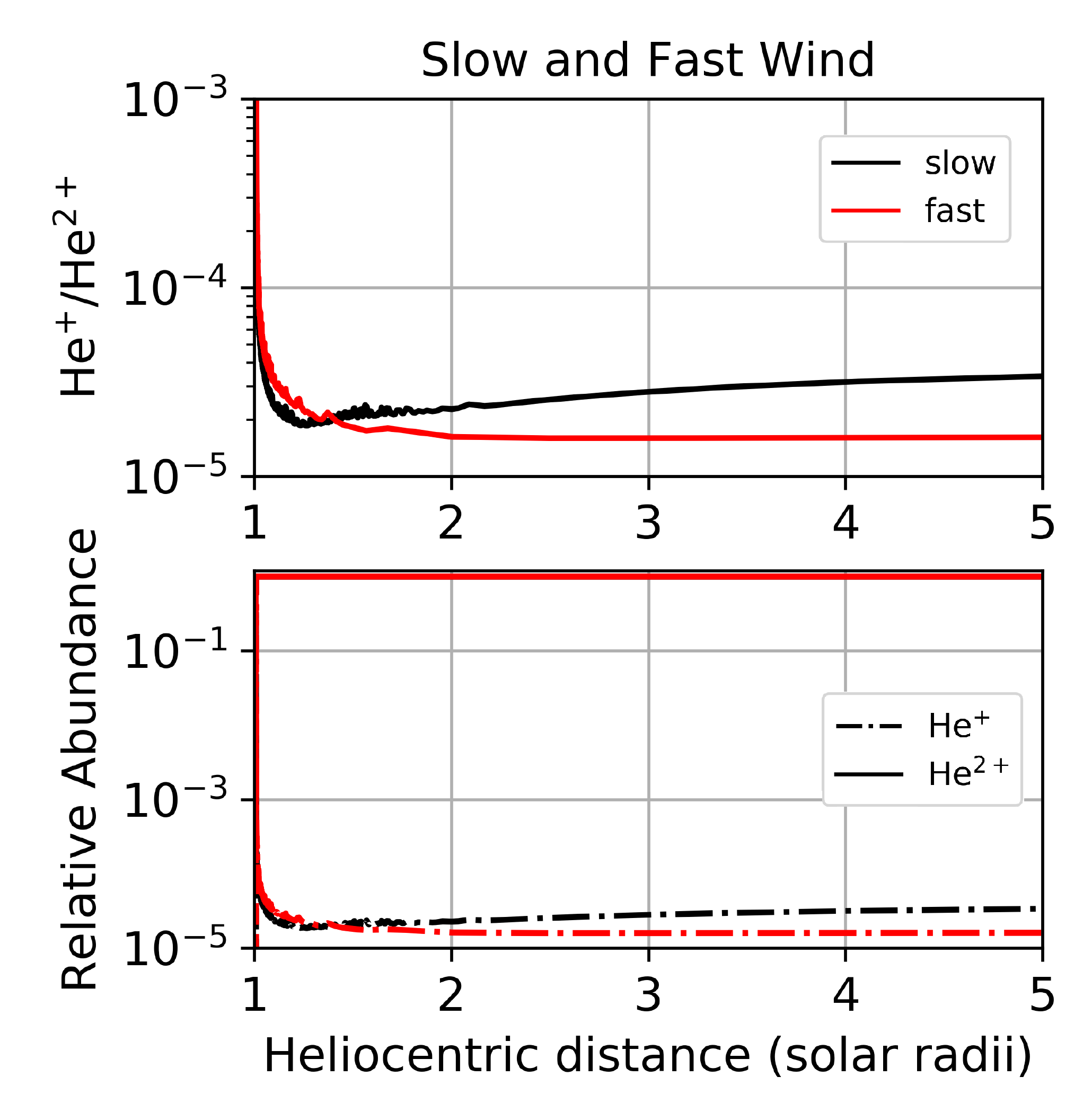}
		\caption{Simulated He evolution within the slow (black) and fast (red) solar wind derived in \cite{Cranmer2007}. Top, density ratio He$^{+}$/He$^{2+}$, and bottom, relative abundance for each ion.}
		\label{fig:IonEvol}
	\end{figure}

	Previously, many non equilibrium ionization modeling studies have aimed to simulate the radial ion evolution of the solar plasma \citep{Rakowski2007, Rakowski2011, Gruesbeck2012, Landi2012, Rivera2019a}. One example is a study using the Michigan Ionization Code (MIC), also used in the present study and further described in Section \ref{meth}, to model the ionization and recombination processes governing the C, O, and Fe charge state evolution within coronal and equatorial solar wind streams \citep{Landi2012}. We use the MIC to simulate the ion evolution of He within the same solar wind of \cite{Cranmer2007} as shown in Figure \ref{fig:SolarWindProfiles}, and present the He evolution in Figure \ref{fig:IonEvol}. The simulations predict a value of He$^{+}$/He$^{2+}$ of 10$^{-5}$ which are $3$ orders of magnitude lower compared to the observations in the present study with a median value of $\sim10^{-2}$. This suggests that either the radial profiles of the solar wind properties do not capture these observations or some other process, unaccounted for in the MIC, is active. 
	
	In a comprehensive study comparing MIC results to measurements of C, N, O, Ne, Mg, Si, S, Fe in the solar wind found that, generally, the three leading solar wind models consistently underestimated the ionization level of all species \citep{Landi2014}. This indicated that the thermodynamic models of the solar wind needed to be adjusted such that it increased the ionization occurring in solar wind plasma. However, increasing the ionization of the plasma is the opposite to what would produce a higher He$^{+}$/He$^{2+}$ necessary to meet observations. This suggests that the ionization codes, rather than the solar wind models, may be missing the necessary process generating the He$^{+}$ that is measured.
	
	We find the He$^{+}$ measurements in the present analysis are, 1) characterized by thermodynamic properties compatible with other solar wind ions, 2) independent of wind type and source region, and 3) do not appear to show a solar cycle dependence. This aligns well with the production of He$^{+}$ from charge exchanges processes that will largely preserve the ion's solar wind identity in phase space, assuming no significant transfer of energy, and are independent of solar wind type. Also, zodiacal light observations measured across the solar cycle are shown to remain stable suggesting the dust, and thus the charge exchange processes and He$^{+}$ densities, are not expected to exhibit a strong solar cycle dependence \citep{Leinert1982}.
	
	We suggest that the He$^{+}$ observed is the solar wind is a signature of the charge exchange process that acts to transform small fraction of alpha particles to He$^{+}$ through this interaction. As mentioned previously, charge exchange between the solar wind and dust neutrals has been investigated in the past (see \citealt{Banks1971, Fahr1981, Gruntman1996}), however the studies did not include observations to constrain modeling results. In the present study, we test charge exchange from several reactions as a possible mechanism in the production of solar He$^{+}$ that is analyzed in Section \ref{ACEObs}.
	
	\begin{table}[t]
		\caption{Charge exchange reactions included in the MIC.}	
		\centering
		\begin{tabular}{c c} 
			\hline\hline 
			\multicolumn{1}{p{2cm}}{\centering Reaction $\#$} &
			\multicolumn{1}{p{5cm}}{\centering Reaction}
			\\
			\hline 
			\\
			1 & $\textbf{He}^{2+} + \textrm{H}_{2} \rightarrow \textbf{He}^{+} + \textrm{H}_{2}^{+}$  \\
			
			2 & $\textbf{He}^{2+} + \textrm{H}_{2} \rightarrow \textbf{He}^{+} + \textrm{H}^{+} + \textrm{H}$   \\
			
			3 & $\textbf{He}^{2+} + \textrm{H}_{2} \rightarrow \textbf{He}^{+} + \textrm{H}^{+} + \textrm{H}^{+} + e$ \\
			
			4 & $\textbf{He}^{2+} + \textrm{H} \rightarrow \textbf{He}^{+} + \textrm{H}^{+}$  \\
			
			\\
			\hline
		\end{tabular}
		\begin{footnotesize}
			Ions in bold originate from the solar wind.
		\end{footnotesize}
		\label{table:xsectreactionrates}	
	\end{table}
	
	\section{Modeling of solar He$^{+}$ ions}\label{meth}

	To quantify the neutral population necessary to produce the He$^{+}$/He$^{2+}$ observed through charge exchange processes, we simulate the radial evolution of He ions with an ionization code that includes the effects of charge exchange with outgassed dust neutrals on a parcel of plasma traveling from the Sun to 1AU. The background neutrals are test particles encountered by the solar wind and their radial distribution remain constant in time while the solar wind alpha particles recombine through charge exchange and thus decrease their abundance.  
		
\begin{figure}[]
		\centering
		\includegraphics[width=0.5\textwidth]{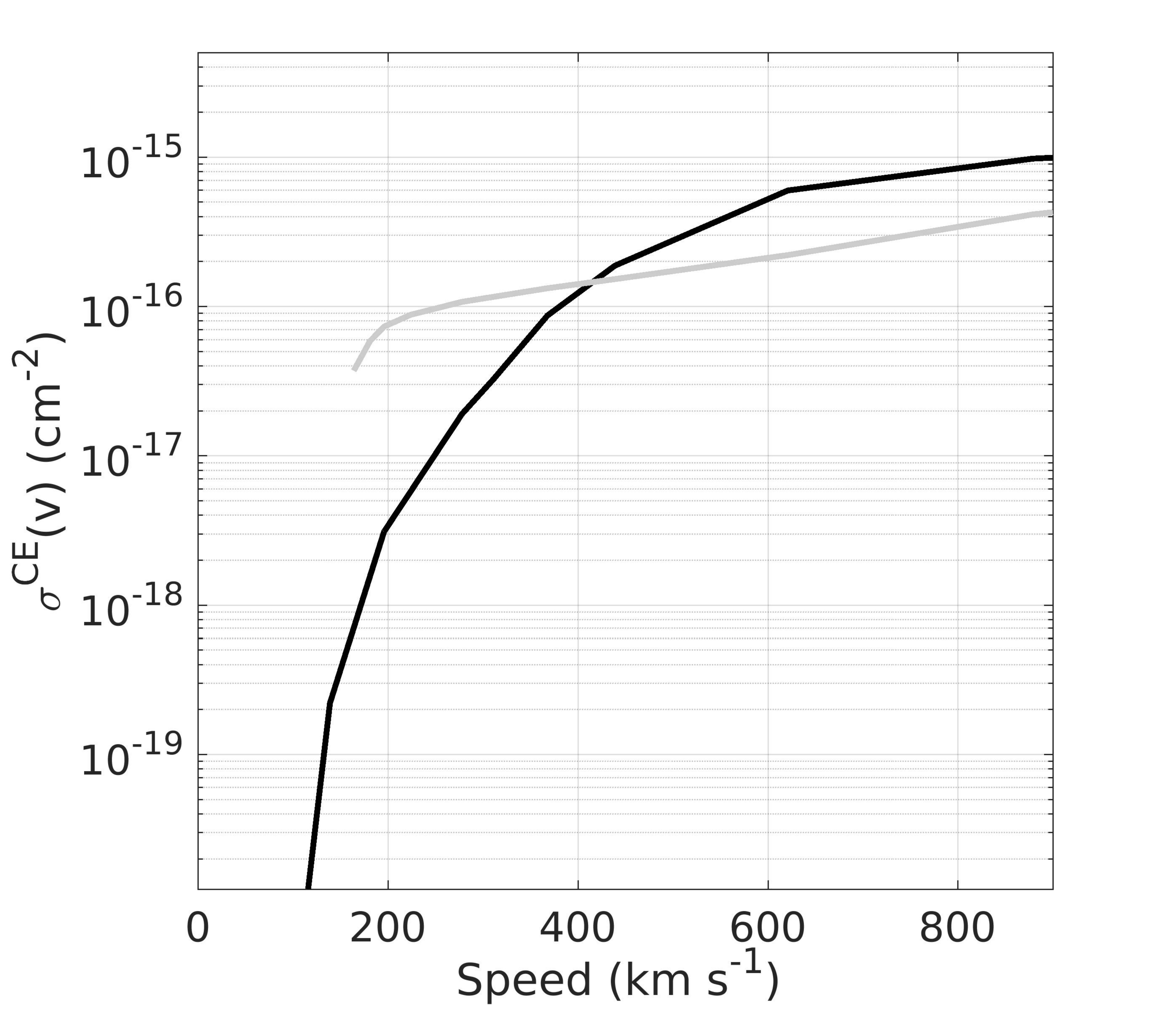}
		\caption{Cross-sections for reaction in Table \ref{table:xsectreactionrates}. The gray curve describes the total cross-section for reaction $1-3$, and the black curve is the cross-section for reaction 4.}
		\label{fig:cross_sects}
	\end{figure}

	We have modified the MIC to simulate alpha particle evolution. The MIC solves a time-dependent set of equations at each step of the plasma's radial expansion using recombination and ionization rates from the CHIANTI 9 atomic database \citep{Dere1997, Dere2019}. The model incorporates the following processes: excitation-autoionization, dielectronic re-combination, collisional ionization, radiative recombination, photoionization, and it has been adapted in the present work to include specific charge exchange reactions from Table \ref{table:xsectreactionrates}. Table \ref{table:xsectreactionrates} includes a collection of reactions previously investigated in the literature that are relevant to the present study \citep{Banks1971, Gruntman1996}. The cross-section values for each reaction are taken from \cite{Barnett1990} and plotted as a function of speed in Figure \ref{fig:cross_sects}.
	
	The charge exchange reactions are included with a term denoted in the equation below as $N_{CE}$,
	
	\begin{eqnarray}
	\frac{dy_{i}}{dt} =  n_{e}\left[y_{i-1}I_{i-1}(T_{e})+y_{i+1}R_{i+1}(T_{e})\right] \nonumber\\ 
	+ y_{i-1}P_{i-1}-y_{i}\left[n_{e}\left(I_{i}(T_{e})+R_{i}(T_{e})\right)+P_{i}\right]
	+ N_{CE} \label{eq1}
	\end{eqnarray}
	
	where $y_{i}$ is the ion's relative abundance of the given element in charge state $i$, $n_{e}$ is the electron density, $T_{e}$ is the electron temperature, $R(T_{e})$ and $I(T_{e})$ are the total recombination and total ionization rates, respectively, that include all mentioned processes except charge exchange. In Equation \ref{eq1}, simulations are for $i=0,1,2$ corresponding to He$^{0,1,2+}$, respectively. The $P$ is the photoionization term described as,
	
	\begin{equation}
	P_{i}=\intop_{\nu_{i}}^{\infty}\frac{4\pi J(\nu)\sigma_{i}(\nu)}{h\nu}d\nu  \label{eq2}
	\end{equation} 
	
	where $J(\nu)$ is the mean spectral radiance of the Sun, $\sigma_i(\nu)$ is the photoionization cross-section for ion $i$, $h$ is the Planck constant, and $\nu_{i}$ is the frequency for the ion's corresponding ionization energy.
	
	The term $N_{CE}$ is defined as follows,
	
	\[N_{CE}= \begin{cases} 
	\Sigma_j (y_2 G_{2,j}^{CE} - y_1 L_{1,j}^{CE}) & \text{ for He}^{+} \\
	\Sigma_j (y_1 L_{1,j}^{CE} - y_2 G_{2,j}^{CE}) & \text{ for He}^{2+} \\
	\end{cases} \tag{8}
	\] \label{eq3} 
	
	where $j$ reactions refer to reactions $1-4$ that are noted in Table \ref{table:xsectreactionrates}. This incorporates the effect of concurrent charge exchange processes with neutrals below 1 AU. $G^{CE}$ in $s^{-1}$ are the charge exchange recombination process and $L^{CE}$ in s$^{-1}$ is the reverse process. For $L^{CE}_{1, j}$, we assume the principle of detail balance such that $L_{1,j}^{CE} = (y_2/y_1)_{TE} G_{2,j}^{CE}$ where $(y_2/y_1)_{TE}$ is the He$^{2+}$/He$^{+}$ charge state ratio in thermal equilibrium. The reaction rates are computed as a function of distance as $G_{2,j}^{CE}(r) = n_{H, H_{2}}(r)v(r)\sigma^{CE}_{2,j}(v)$ where $n_{H, H_{2}}(r)$ is the number density of H or H$_{2}$ with heliocentric distance, $r$, from the Sun. The $\sigma^{CE}_{2,j}(v)$ is the cross-section for each charge exchange process, $j$, between He$^{2+}$ and neutrals.
	
	\begin{figure*}[]
	\centering
	\includegraphics[width=0.7\textwidth]{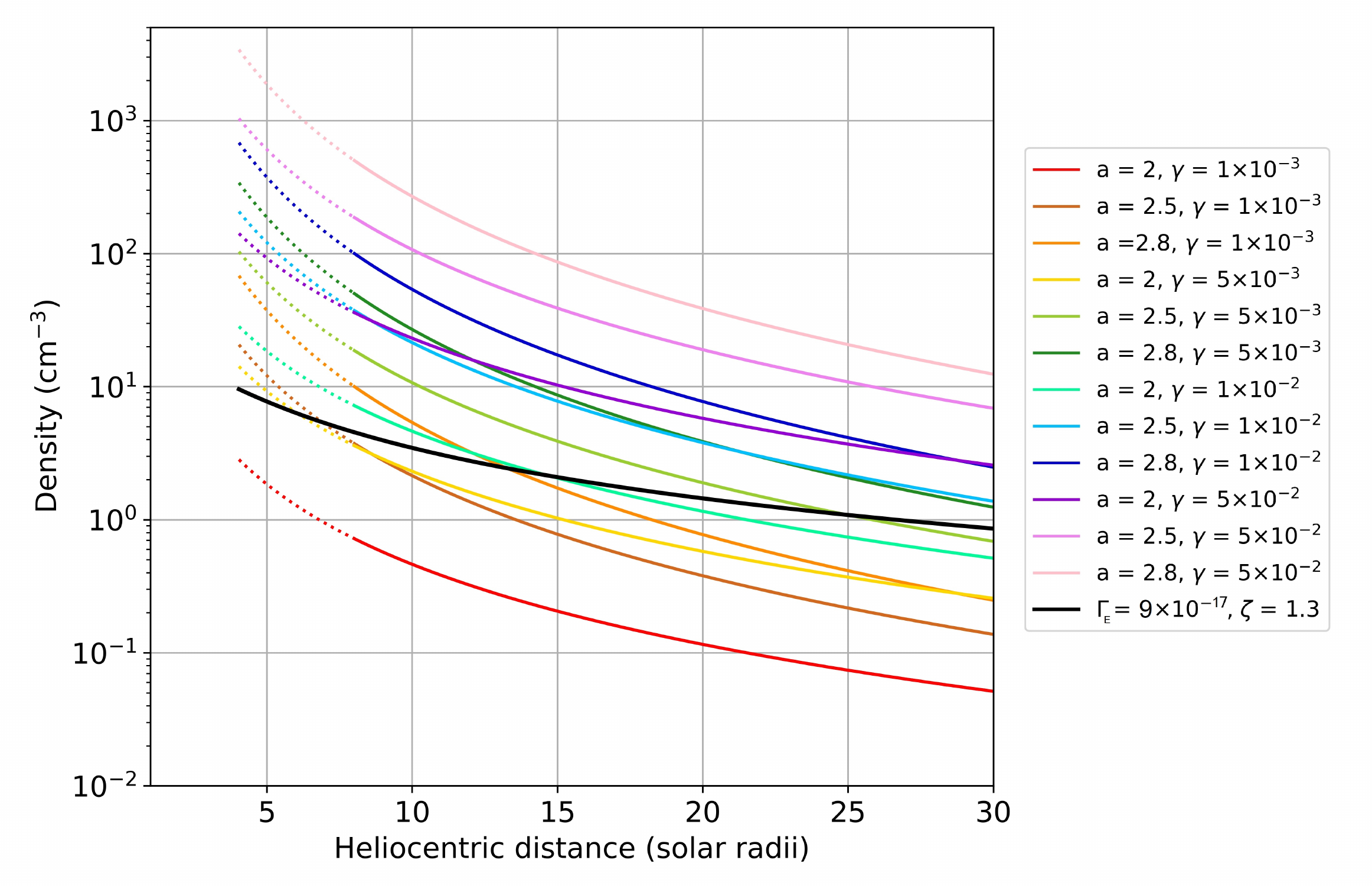}
	\caption{Empirical models of $\text{H}+\text{H}_{2}$ density profiles with an inner edge of $4R_{\Sun}$ starting at the dotted line and $8R_{\Sun}$ starting at the solid line. The solid line is the density of $n_{H+H_2}$ predicted using the dust geometric factor, $\Gamma=\Gamma_{E}(r_E/r)^\zeta$, see description in Section \ref{discussion}.}
	\label{fig:dustdens}
\end{figure*}

\begin{figure*}[t]
	\centering
	\includegraphics[width=\textwidth]{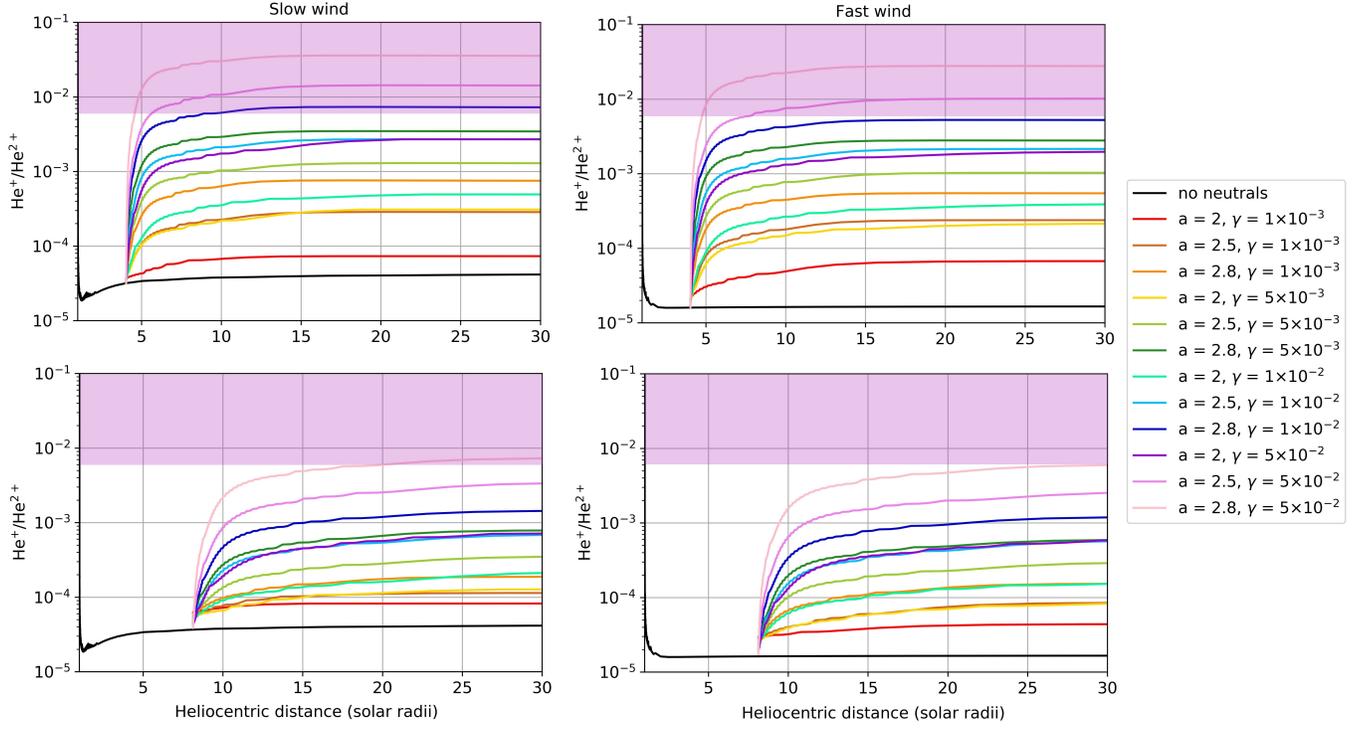}
	\caption{He charge state evolution generated using properties of slow (left column) and fast (right column) solar wind between 1 to $30R_{\Sun}$. The plots show the radial evolution of He$^{+}$/He$^{2+}$ for a neutral profile with an inner boundary of $4R_{\Sun}$ (top) and $8R_{\Sun}$ (bottom). The shaded purple region covers the range of He$^{+}$/He$^{2+}$ values found in the observations.}
	\label{fig:HeEvolRsun}
\end{figure*}

	The dust neutrals within the interplanetary medium are test particles in the simulation. We assume a static profile for the population of outgassed H and H$_{2}$ originating from the circumsolar dust grains. The dust has been observed to be distributed symmetrically across the ecliptic plane with a large concentration in the vicinity of the Sun. The outgassed neutral profile in our simulations assumes a symmetric and constant distribution surrounding the Sun that coincides with the dust population. This assumption holds given that the dust structure is not observed to undergo large changes within the timeframe of the solar wind propagation to 1 AU, which is on the timescale of days. 
	
	The neutral dust profile, $n$ (cm$^{-3}$), is taken to be in the form, 
	
	\[n_{H+H_{2}}(r)= \begin{cases} 
	0 & r < \text{Inner edge } (AU) \\
	\gamma\cdot r^{-a} & r > \text{Inner edge } (AU) \\
	\end{cases} \tag{9}
	\] \label{neutralprofile} 
	
	where $\alpha$ and $\gamma$ are free parameters that govern the density of neutral dust material available for charge exchange. The radial distance, $r$, is in AU. The density profiles and the corresponding parameters used are shown in Figure \ref{fig:dustdens} where the neutral profiles are a sum of parts H and H$_{2}$ where $n_{H_2}=8\cdot n_{H}$ as is approximated to be outgassed by the dust in \cite{Gruntman1996}.	In addition, the solid black curve is the $n_{H+H_{2}}$ predicted by previous models, further discussed in Section \ref{discussion}.
	
	Many different combinations of the free parameters have been used to test a range that may produce the observed He$^{+}$/He$^{2+}$ measured at 1 AU. Parameters are motivated by observations of the distribution of dust from zodiacal light and F-corona studies previously found in the literature \citep{Kimura1998, Kimura1998a, Mann2004}. There is still a large variation between the distribution of interplanetary dust inferred through remote sensing measurements. Observations of the F-corona continuum in the near-infrared show a sharp drop off in brightness that scale between r$^{-1.9}$ to r$^{-2.5}$ for the equatorial and r$^{-2.3}$ to r$^{-2.8}$ for the polar solar regions below 10$R_{\Sun}$ which correspond to freeze-in distances in our simulations \citep{Koutchmy1985, MacQueen1995, Leinert1998}. In \cite{Gruntman1994, Gruntman1996}, both $n_{H}$ and $n_{H_2}$ were estimated as $\propto r^{-1.15}$ below 1AU. However neutrals in the heliosphere are highly uncertain and remain relatively unknown in the vicinity of the Sun. For this reason, we explore a different range of values.
	
	Furthermore, near-Infrared measurements observe an intensity enhancement at an elongation angle, $e$, of $1^\circ$.  The elongation angle is defined as the angle made between the line of sight of the observer and the Sun's center. The distance from the Sun along the plane of the sky for a corresponding $e$ is $r = \sin(e)\cdot 1$(AU). For $e=1^\circ$, this equates to approximately $r=4R_{\Sun}$ on the plane of the sky. An elevated intensity in remote measurements at this distance has long suggested a possible inner edge of the dust at that location \citep{Kimura1998, Kimura1998a, Mann2004}. The empirical neutral profile we test assumes a theoretical inner edge that begins at $4R_{\Sun}$. In addition, we investigate distributions with an inner edge as far out as $8R_{\sun}$ to evaluate the sensitivity of the results to the location of the inner edge.
	
	To simulate the solar wind ion evolution, we use models for the equatorial (slow) and coronal hole (fast) solar wind derived from \cite{Cranmer2007} shown in Figure \ref{fig:SolarWindProfiles}. 
	
	\section{Modeling Results} \label{results}
	We simulated the charge state distribution of He as a function of distance using the modified MIC that includes charge exchange with the surrounding outgassed dust neutrals. For each neutral distribution in Figure \ref{fig:dustdens}, we present the corresponding simulated He$^{+}$/He$^{2+}$ radial evolution shown in the top plot of Figure \ref{fig:HeEvolRsun}, for the slow (left) and fast (right) solar wind. In addition, we include the solutions for a neutral inner edge of $8R_{\Sun}$ in the bottom of Figure \ref{fig:HeEvolRsun} where the simulated He charge states are presented in the same manner. Each plot includes the range of observed He$^{+}$/He$^{2+}$ observations denoted as the shaded purple region. 
	
	We test values between a power law exponent of $a = 2.0$ and $a = 2.8$ with increasing $\gamma$ until the He$^{+}$/He$^{2+}$ value reaches the same order of magnitude as the mean He$^{+}$/He$^{2+}$ observed with ACE/SWICS, $\sim10^{-2}$. The results show a sharp increase of He$^{+}$/He$^{2+}$ as the simulated plasma parcel encounters the dust. This is due to the step function used to describe the dust in the inner corona; however, the region below the inner edge we define could be decreasing to zero in a smoother way.

	In both the slow and fast cases, we can begin to see a deviation between the He values with the presence of neutrals immediately at the inner edge of the dust. The slow and the fast solar wind values respond similarly to charge exchange processes even though their thermodynamic properties can differ significantly during the radial expansion. This is consistent with the He$^{+}$/He$^{2+}$ observations which are found to be largely independent of solar wind type e.g. solar wind speed and composition. 
	
	\section{Discussion} \label{discussion}
	We find the He$^{+}$/He$^{2+}$ simulated values are sensitive to the density for all the neutral profiles tested. The neutral profiles that generate a value of He$^{+}$/He$^{2+}$ within the range of the observations correspond to $a = 2.8$ and $\gamma>1\times10^{-2}$, $a=2.5$ and $\gamma>5\times10^{-2}$, for an inner boundary of $4R_{\Sun}$, shown at the top of Figure \ref{fig:HeEvolRsun}. For an inner boundary of $8R_{\Sun}$, shown at the bottom of the same figure, the neutral profile within the observational range is $a = 2.8$ and $\gamma > 5\times10^{-2}$.
	
	In addition, the He values are sensitive to the location of the inner edge along with the density of the dust profile. A distribution with an inner edge closer to the Sun produces an overall larger He$^{+}$/He$^{2+}$ heliospheric value when maintaining the same $a$ and $\gamma$ values. Comparing the rows of Slow and Fast wind, the profile with parameters, $a=2.8$ and $\gamma = 5\times10^{-3}$, with an inner edge of $4R_{\Sun}$ produced a He$^{+}$/He$^{2+}$ value roughly an order of magnitude higher compared to a profile with an inner edge of $8R_{\Sun}$. This may be a result of charge exchange occurring earlier in the evolution, giving the alpha particles more time to recombine prior to the freeze-in height of He$^{2+}$. Alternatively, given that the reaction rates, L and G, depend on the neutral density, a higher density of neutral material can also contribute to a larger He$^{+}$/He$^{2+}$ ratio overall.
	
	Moreover, we expect the majority of neutral H and H$_{2}$ closest to the Sun to originate from dust while the interstellar neutral values begin to dominate much farther from the Sun \citep{Fahr1981}. From Figure \ref{fig:HeEvolRsun}, we find the ions are highly sensitive to the presence of neutrals below $15R_{\Sun}$, for the slow, and $30R_{\Sun}$, for the fast, where the dust component dominates. Therefore, we expect that the majority of solar He$^{+}$ predicted to be formed through charge exchange from outgassed dust neutrals rather than interstellar material.
	
	We compare our results to previous values of outgassed H$_{2}$ and H densities \citep{Fahr1981}. Following \cite{Fahr1981}, the production rate for H and H$_{2}$ is given as, P$_{D}(r)=n_{p}(r)v_{rel}(r)\epsilon\Gamma(r)$, where $n_{p}(r)$ is the proton density with radial distance, $r$, $v_{rel}(r)$ is the relative velocity between solar wind and the dust grains, $\epsilon=0.9$ and $\epsilon=0.05$ is the desorption efficiency for H$_{2}$ and H, respectively, $\Gamma(r)$ in cm$^{-1}$ is the dust cross sectional area per unit volume of the dust,
		
	\[
		\Gamma(r) =\int_{s_{min}}^{s_{max}} \pi s^2 f(s,r)ds
		\tag{10}
	\]
		
	 where $s$ is the radius of the grain, $f(s,r)$ is the distribution of grains with grain size, $s$, and distance from the Sun, $r$.
	 	
	 Following equilibrium conditions, the P$_{D}$ equates to the destruction, D$_{D}(r)$, of H and H$_{2}$. D$_{D}(r) = n_{H, H_2}(r)(C^{p}(r)+C^{e}(r)+C^{exc}(r))$ where $C^{p}(r)$ is the photoionization rate, $C^{e}(r)$ is the electron impact ionization rate, and $C^{exc}(r)$ is the proton charge exchange rate of H and H$_{2}$ with distance from the Sun. Rearranging, density can be computed as the sum of $n_{H}(r)$ + $n_{H_{2}}(r)$ in cm$^{-3}$ as the following,
	
		\[
		n_{l}(r) = \Sigma_{l}\frac{n_{p}(r)v_{rel}(r)\epsilon_{l}\Gamma(r)}{C^{p}_{l}(r)+C^{e}_{l}(r)+C^{exc}_{l}(r)} \tag{11}
		\]
	
	for $l = $ H$_{2}$ and H. Assuming, $C^{p}(r) = C^{p}_E(r_E/r)^2$, $C^{e}(r) = C^{e}_E(r_E/r)^2$, and  $C^{exp}(r) = C^{exp}_E(r_E/r)^2$ where the subscript 'E' refers to that value at 1AU. $C^{p}_E = 3.67\times10^{-7}, 1\times10^{-7}$ s$^{-1}$, $C^{e}_E = 1\times10^{-7}, 1\times10^{-7}$ s$^{-1}$, and $C^{exc}_E = 1\times10^{-8}, 1\times10^{-7}$ s$^{-1}$ for H$_{2}$ and H, respectively \citep{Banks1971, Gruntman1996}. There is still large uncertainty on the value of $\Gamma$ and how it changes with radial distance from the Sun. Previous values of $\Gamma$ span five orders of magnitude, $\sim10^{-17}-10^{-21}$ \citep{Bame1968,Banks1971, Holzer1977}. \cite{Fahr1981} approximated the geometric factor as $\Gamma(r)=\Gamma_{E}(r_{E}/r)^{\zeta}$ where $\Gamma_{E}=2\times10^{-19}$ cm$^{-1}$ and $\zeta=1.3$. At $4R_{\Sun}$, this produces $n_{H+H_2}\sim5\times10^{-2}$ cm$^{-3}$ which is 4 orders of magnitude lower compared to the minimum density that would generate He$^{+}$/He$^{2+}$ to meet observational values for the $4R_{\Sun}$ profile case, $a=2.8$ and $\gamma=1\times10^{-2}$. For an upper limit value of $\Gamma = 9\times10^{-17}$, the density predicted increases to  $n_{H+H_2}\sim10$ cm$^{-3}$ at $4R_{\Sun}$ which is two to three orders of magnitude lower compared to our required value, as shown with the solid black curve in Figure \ref{fig:dustdens}. Generally, our results suggest a much larger dust density profile or distribution of grain size, as described by $\Gamma(r)=\Gamma_{E}(r_{E}/r)^{\zeta}$, than previously predicted in the vicinity of the Sun to generate the H and H$_{2}$ densities necessary to meet observational values.
		
	Our results may also suggest that there could be an additional process contributing to the formation of solar-like He$^{+}$. One possibility would be the ionization of neutral He outgassed from the dust. This process would produce He$^{+}$ where the dust is present that could potentially have enough time to thermalize by 1AU. If so, this could generate a non-solar He$^{+}$ VDF with a Maxwellian-like profile rather than a typical PUI profile. This scenario would require further testing with the MIC, along with kinetic modeling of the VDF evolution from its creation in the corona to 1AU. 
	
	Moreover, our simulations predict the presence of neutral material in the extended corona which coincides with several decades of eclipse observations which observe He I 10830 in the vicinity of the Sun. The diffuse neutral helium found in eclipse observations was initially attributed to being geocoronal and to interstellar material \citep{Kuhn1996,Kuhn2007}; however, the most recent observations have linked the neutral He to the Sun \citep{Moise2010, Dima2018}. These studies suggest the diffuse He signal arises from the interaction between the solar wind alphas and surrounding dust, essentially acting to neutralize the alpha particles as they propagate from the Sun. In a similar manner to the present study, we can independently determine the density of neutral material in the extended corona by focusing on charge exchange reactions between H and H$_{2}$ that can neutralize alpha particles. This can be another method of further constraining neutrals near the Sun.
	
	Furthermore, if the production of solar He$^+$ is due to the presence of the dust, then there will likely be a reduction in He$^{+}$/He$^{2+}$ at higher latitudes where dust is depleted compared to the ecliptic.  This can be achieved with future off ecliptic measurements from the Heavy Ion Sensor (HIS) on Solar Orbiter \citep{Muller2020}. The Ulysses spacecraft has previously sampled the solar wind above and below the ecliptic outside of the orbit of the Earth, however HIS will do this around the orbit of Mercury ($\sim60R_{\Sun}$) where we expect the majority of newly generated He$^{+}$ to be primarily formed through dust interaction in the inner heliosphere.
	
	Another important observational constraint to the charge exchange process is the emission of X-ray and far-ultraviolet radiation. Observations from the coma of Comet C/1996 B2 Hyakutake were discovered to emit a strong X-ray and EUV signal as a byproduct of charge exchange between neutral cometary material and solar wind ions \citep{Lisse1996,Cravens1997}. From laboratory experiments, photon emission through electron capture show that solar wind $\alpha$ particles and H and H$_{2}$ from interplanetary dust should produce emission in the far-ultraviolet (FUV) or EUV spectral range from He II 304\AA, for He$^{+}$ produced, and He I 584\AA, for neutralized He formed \citep{Bodewits2006}.
	
	Lastly, in our simulations, we specified a static neutral distribution; however the dust can vary at different spatial scales. Solar transients and periodic deposits of fragmented comet material near the Sun can disrupt the homogeneity of the dust \citep{Jones2018}. This can develop local temporal and spatial changes in the dust allowing for periodic enhancements or depletions in neutral density that could result in sporadic fluctuations of He$^{+}$/He$^{2+}$ values. This could explain the relatively large fluctuations that are occasionally observed, as shown in Figure \ref{fig:HeliumTime}.
		
	\section{Summary and Conclusions} \label{conclusion}
	
	In the present work, we investigate the presence of He$^+$ observed by ACE/SWICS in the solar wind throughout the majority of cycle 23 ($1998-2011$) in order to understand the origin and mechanism that produce this ion. From the ACE/SWICS dataset, we identified several periods where He$^{+}$ VDFs can be well modeled by a Maxwellian distribution traveling near the corresponding solar wind proton speed whose temperature follows a super mass proportional relationship that coincides with previous studies of ions in the solar wind. These properties suggest the He$^{+}$ ions resemble material of solar origin.
	
	Furthermore, He$^+$ does not depend on the solar cycle, wind speed, composition, and source region, suggesting that the mechanism producing He$^{+}$ is independent of processes that generate the distinctive properties of the solar wind but rather a process common to wind traveling on the ecliptic and occurring before reaching 1AU. 
	
 	Current simulations of He charge state distributions accounting for electron impact ionization and recombination along with photoionization and radiative recombination processes are shown to underestimate He$^{+}$/He$^{2+}$ by $3-4$ orders of magnitude. This indicated the possibility of missing processes that may be active during the radial expansion of the solar wind. To reconcile the missing He$^+$, we tested the effectiveness of charge exchange between the solar wind and H and H$_{2}$ from dust in the production of He$^{+}$ to determine if this recombination process could explain the enhancement of He$^+$/He$^{2+}$ at 1AU. Our work estimates the radial distribution and density of neutrals that are required to be present during the solar wind's radial expansion which could make charge exchange an effective method of transforming a small fraction of solar alpha particles into singly ionized He to meet observational values. Results show that a distribution of dust neutrals (H, H$_{2}$) following a power law of the form $\gamma r^{-a}$, with $\gamma > 0.05$ for $a = 2.8$, with an inner edge between $4-8R_{\Sun}$ can produce the  He$^+$/He$^{2+}$ values. Our results predict a density of H and H$_{2}$ that is several orders of magnitude larger than previous models, suggesting; 1) that dust density and distribution of grain size in the vicinity of the Sun is larger than previously described, and/or 2) there may be additional processes contributing to the solar He$^{+}$ population below 1AU.

	Further observations are needed to better constrain the density of H and H$_{2}$ near the Sun. However, our modeling results may be an indication of the importance in accounting for processes between the dust and solar wind. One of the key implications from our results is the importance of charge exchange processes during the radial propagation of the solar wind. This may be a significant process shaping the ion composition measured in the heliosphere. 
	
	Further constraints to the presence of neutrals are necessary to continue testing our hypothesis more rigorously. Remote observations remain limited to the progress of dust studies due to the loss of spatial information from intensities integrated along the line of sight. A more effective manner of probing the dust environment near the Sun is through a combination of remote and in situ observations. 
	
	\acknowledgements
	The authors thank the anonymous referee for insightful comments that greatly improved the final manuscript. Y. J. Rivera was supported by a Rackham Graduate School Fellowship from the University of Michigan and the Newkirk Fellowship from the High Altitude Observatory. Y. J. Rivera thanks Jim Raines, Sarah Spitzer, and Ryan Dewey for valuable discussion in the preparation of this manuscript. EL acknowledges support from NASA grants 80NSSC18K0647, 80NSSC18K1553, 80NSSC20K0185 and NSF grants AGS-1408789 and AGS-1621686. S. T. Lepri was supported by NSF grant award AGS-1460170, AGS-1358268  NASA grants 80NSSC18K0645, 80NSSC20K0185, 80NSSC19K0853, 80NSSC20K0192, and 80NSSC18K0101. Y. J. Rivera and S. T. Lepri acknowledge the International Space Science Institute in Bern, Switzerland, for helpful discussion related to this work. CHIANTI is a collaborative project involving George Mason University, the University of Michigan (USA) and the University of Cambridge (UK).

\end{document}